\documentclass[journal,10pt,final]{IEEEtran}

\usepackage{amsmath}
\usepackage{graphicx}
\usepackage{amssymb}
\usepackage{multirow}

\usepackage{xcolor,color}
\definecolor{mirojo}{rgb}{1,0,0}
\definecolor{mverde}{rgb}{0,1,0}
\definecolor{mazul}{rgb}{0,0,1}
\definecolor{mmag}{rgb}{1,0,1}

\usepackage{tikz}
\usetikzlibrary{arrows, calc, fit, shapes, positioning}

\begin{document}

\title{Direct Polarization for q-ary Source and Channel Coding }

%%importante: el thanks tiene que estar dentro del autor. En caso
%%contrario aparece una página en blanco.
\author{ \'Angel Bravo-Santos${}^*$,~\IEEEmembership{Senior Member,~IEEE,}%
  \thanks{This work has been partially supported by the Spanish government (TEC2012-38883-C02-01, Consolider-Ingenio 2010 CSD2008-00010).}%
  \thanks{A. Bravo-Santos is with the Departamento de Teor\'{\i}a de la
    Se\~nal y Comunicaciones, Universidad Carlos III de Madrid,
    Avda. de la Universidad 30, Legan\'es, 28911 Madrid, Spain
    (abravo@tsc.uc3m.es), {\it Tel:} +34 91 624 8752, {\it Fax:} +34
    91 624 8749.}}%
\maketitle
\IEEEpeerreviewmaketitle

\begin{abstract}
  It has been shown that an extension of the basic binary polar
  transformation also polarizes over finite fields. With it the direct
  encoding of q-ary sources and channels is a process that can be
  implemented with simple and efficient algorithms. However, direct
  polar decoding of q-ary sources and channels is more involved. In
  this paper we obtain a recursive equation for the likelihood ratio
  expressed as a LR vector. With it successive cancellation (SC)
  decoding is applied in a straightforward way. The complexity is
  quadratic in the order of the field, but the use of the LR vector
  introduces factors that soften that complexity. We also show that
  operations can be parallelized in the decoder.  The Bhattacharyya
  parameters are expressed as a function of the LR vectors, as in the
  binary case, simplifying the construction of the codes.  We have
  applied direct polar coding to several sources and channels and we
  have compared it with other multilevel strategies. The direct q-ary
  polar coding is closer to the theoretical limit than other
  techniques when the alphabet size is large. Our results suggest that
  direct q-ary polar coding could be used in real scenarios.
\end{abstract}


\section{Introduction}

The basic polar transformation
for channel coding presented in \cite{arikan09} maintains the polar
properties for certain alphabets and channels.  The authors in
\cite{sasoglu09} prove that, under certain conditions, the mutual
information of q-ary channels polarizes. Moreover, if the input
alphabet is of prime size with a modulo addition defined in it, the
basic transformation polarizes. In the same paper it is also shown
that for some non-prime alphabets it is possible to find channels for
which the basic transform does not polarize.

In \cite{sahebi2011multilevel} and \cite{park2013polar} it is shown
that for alphabets whose size is a power of a prime the basic polar
transformation for channel coding polarizes with more than two
levels. Specific definitions of Bhattacharyya parameters allow the
construction of informations sets that, eventually, allow for reliable
transmission of information. Further, the multilevel channel polar coding
reaches the symmetric capacity.

Polar codes have been applied to continuous channels using some form
of multilevel modulation. In \cite{trifonov12} generalized
concatenation codes implemented using polar codes are applied to pulse
amplitude modulation (PAM). In \cite{seidl13} a
multilevel code is defined as a binary
partition of the multilevel channel that is concatenated with a polar
code. It is also shown in \cite{seidl13} that this scheme
reaches the capacity of the symmetric channel.

Binary polar codes have been used for lossy
source coding using distortion criteria \cite{korada09}. In \cite{hussami09} lossless
source coding was described based on the dual channel approach. A more
direct approach to source coding with binary polar codes was provided
by Arikan in \cite{arikan10}, where it was also shown that channel
coding can be interpreted as an instance of source coding.

The extension of  binary polar codes for source coding to the q-ary
case under a distortion criterion
was presented in \cite{karzand10}. The work \cite{sasoglu11}
generalizes lossless polar source coding to arbitrary alphabets using
the Arikan's direct source coding approach \cite{arikan10}. In
\cite{sasoglu11} it was also proved that the basic
polar transform \cite{arikan09} always polarize with alphabets of prime
size, but there are alphabets of non-prime size where it is possible
to find channels where the polarization is not possible. In the same paper
the channel coding problem is treated as an instance of source coding. In
\cite{mori14} it is shown that the basic transformation can be generalized
allowing polarization of sources and channels in finite fields.

There is a well known optimal symbol-by-symbol decoding rule in the
literature \cite{hartmann76} for linear codes with symbols defined on
a finite field. Its complexity is exponential in the dimension of the
dual code and therefore more efficient alternatives are used. The usual way of implementing ``good'' codes with q-ary
alphabets, or q-ary signal sets, is by means of constructing first ``good'' binary codes,
where ``good'' refers to the proximity to the theoretical limit \cite{wachsmann99}. The
main objective of this work is to find the theoretical tools to
implement and construct polar codes that work natively with q-ary
alphabets, or q-ary signal sets. Therefore this direct approach, that
works with symbols from a finite field and with a polar transformation
also defined in that field \cite{arikan15}, avoids the use of binary
codes. In previous works \cite{sasoglu11,mori14}, no explicit
expression or algorithm was provided for decoding. 

In this paper we use the LR vector %
of the encoded symbols given the side information for source coding,
or the LR vector of the unencoded symbols given the channel
observations, for non-binary polar codes. We then derive a recursive expression for the LR
vector that allows to perform successive cancellation (SC) decoding.
The Bhattacharyya parameters are obtained as a function of the LR. As
a result, the problems of q-ary source and channel coding, and code
construction, are solved as in \cite{arikan09} with complexity of the
same order as in the binary case in the encoder and about $q^2/2$
times the binary complexity in the decoder.

In this paper we also perform a comparison of q-ary polar coding for
channel coding with multilevel coding (MLC) and bit-interleaved coded
modulation (BICM) using simulations. The results suggest that q-ary
polar codes are closer to the theoretical limit that other codes as
the alphabet size is large. There is not need to optimize the labeling
symbols-signals in channel coding with q-ary polar codes. However,
this is not the case of other multilevel coding
alternatives. Furthermore, the relation between channel degradation
and Bhattacharyya parameters of the input symbols in q-ary polar
coding is similar to the binary case, as it is shown in the
paper. From this fact it is straight forward to extend the benefits of
binary polar coding to the q-ary case, such as, rate adaptation
\cite{bravo13ray}, network coding \cite{andersson10,bravo13rel},
etc. This extension is not so immediate with MLC or BICM. The complexity of q-ary polar decoding is higher than that of MLC or
BICM. However, as it is discussed in the paper, many operations can be
done in parallel, as opposed to the serial definition of MLC.

In this work we follow the notation in \cite{arikan09} and we do not
provide additional explanation about it. The rest of the paper is
organized as follows. In Section II we formulate the problem. In
Section III we present the recursive LR equations and Bhattacharyya
parameter as a function of the LR. In Section IV we analyze the
implementation of the source and channel polar coding. In Section V we
describe the simulations conducted with q-ary sources and multilevel
polar coding on the additive white Gaussian noise (AWGN) channel. The
main conclusions about the results of this work are in Section VI.
 
\section{Source Coding with Polar Codes over Finite Fields
\label{sec:problem-statement}
}

\begin{figure}[h]
  \centering
  \includegraphics[width=8.5cm]{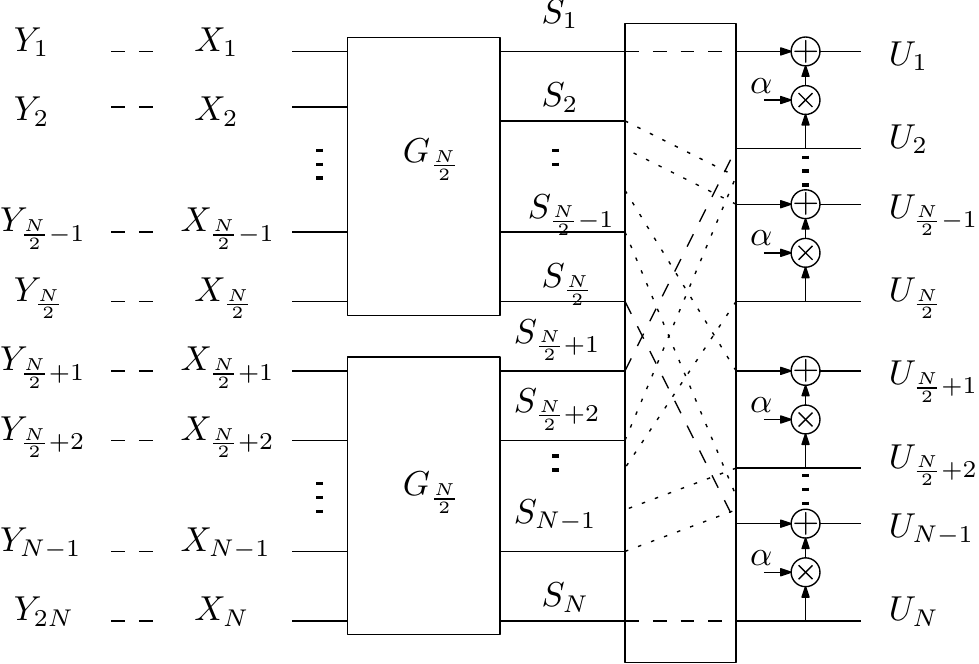}
  \caption{q-ary polar encoder.}
\label{fig:1}
\end{figure}

We consider  a discrete memoryless source with alphabet $\cal X$ formed by the
elements of the Galois field $\mathbb{F}_q$. Associated to a source symbol
$X$ is the side information $Y\in{\cal Y}$. We have $(X,Y)\in {\cal
  X\times Y}$. The pair $(X,Y)$ can be view as a memoryless source,
$(X,Y)\sim P_{X,Y}(X,Y)$, where $P_{X,Y}(X,Y)$ is the
joint distribution of $(X,Y)$ \cite{arikan10}. By taking symbols independly from this
source we form the sequence 
$$(X_1^N,Y_1^N)=\big( (X_1,Y_1), (X_2,Y_2)\cdots
(X_N,Y_N)\big),\ N=2^n,\ n\in\mathbb{N}.$$
We compress the source sequence $X_1^N$ using a polar code. 
The basic polar transformation was introduced in \cite{arikan09} and  extended in \cite{mori14} for $\mathbb{F}_q$. The extended transformation is,
\begin{equation}
  \label{eq:1}
  \begin{aligned}
    U_1&=X_1+\alpha X_2\\
U_2&=X_2,
  \end{aligned}
\end{equation}
where %
$U_1, U_2, X_1, X_2\in \mathbb{F}_q$, the addition is in
$\mathbb{F}_q$, and $\alpha$ is a constant that is a primitive element
of the field if $\mathbb{F}_q$ is a non-prime field. If the field is prime then
$\alpha=1$. %
The transformation \eqref{eq:1} polarizes over $\mathbb{F}_q$, $q=p^m$,
$m,p\in\mathbb{N}$, $m>1$, $p$ prime
\cite{mori14}, or over $\mathbb{F}_p$  \cite{sasoglu09,sasoglu11}.
% and \eqref{eq:1} is the basic transformation that polarizes
%$\mathbb{F}_p$, $p\in \mathbb{N}$, $p$ prime
%\cite{sasoglu09,sasoglu11}.  In \cite{mori14} it was proved that the
%extended transformation polarizes for symbols
%$U_1, U_2, X_1, X_2\in \mathbb{F}_q$.
Based on \eqref{eq:1} the encoder for the sequence $(X_1^N, Y_1^N)$ follows the classical scheme for polar channel coding
described in \cite{arikan09}. In Fig. \ref{fig:1} we show the
structure implemented in the encoder. It is easy to expresses the
transformation of Fig. \ref{fig:1} as
\begin{equation}
  \label{eq:1b}
  U_1^{N}=X_1^{N} G_{N},
\end{equation}
where $G_{N}$ is the matrix of the transformation.
%In polar source coding the element $U_i$ given $(Y_1^{N},U_1^{i-1})$ plays a special role. We call it coordinate
%element. This definition is parallel to that in \cite{arikan09} for channels.   

In \cite{sasoglu2012polar} and \cite{mori14} it is proved that the
entropy of $U_i$ given $(Y_1^{N},U_1^{i-1})$ polarizes on
$\mathbb{F}_p$ with the transformation \eqref{eq:1}. A similar result was proved in
\cite{mori14} for $\mathbb{F}_q$. For $\epsilon>0$
\begin{equation}
  \label{eq:1c}
  \begin{gathered}
    \lim_{N\to \infty}\frac{1}{N}\left|\{i:H(U_i|Y_1^N U_1^{i-1})>1-\epsilon \}\right|=H(X|Y),\\
    \lim_{N\to \infty}\frac{1}{N}\left|\{i:H(U_i|Y_1^N U_1^{i-1})<\epsilon \}\right|=1-H(X|Y),\\
  \end{gathered}
\end{equation}
where the logarithms are to base $q$. A conclusion
from \eqref{eq:1c} is that if N is large enough we only have two
possibilities related to the amount of information about $U_i$ given $(Y_1^{N},U_1^{i-1})$: a) if $H(U_i|Y_1^N, U_1^{i-1})\approx
0$ the knowledge of $(Y_1^N, U_1^{i-1})$ is all we need for recovering
$U_i$; b) if $H(U_i|Y_1^N, U_1^{i-1})\approx
1$, the pair  $(Y_1^N, U_1^{i-1})$ does not give information about $U_i$.

In \cite{arikan10} Arikan gave a definition of the Bhattacharyya parameter of a binary source. This definition was extended in \cite{sasoglu2012polar} for non-binary alphabets. For a generic source $(X,Y)$, $X\in {\cal X}$, the Bhattacharyya parameter is
\begin{equation}
  \label{eq:1d}
  Z(X|Y)=\frac{1}{|{\cal X}| -1}\sum_{\substack{x,x^\prime\in {\cal
        X}\\ x\ne x^\prime}}\sum_y\sqrt{P_{X,Y}(x,y) P_{X,Y}(x^\prime, y)}.
\end{equation}
For the variables $(U_i; Y_1^N,U_1^{i-1})$, $U_i\in \mathbb{F}_q$, we have,
\begin{equation}
  \begin{aligned}
    Z(U_i| &Y_1^N,U_1^{i-1})=\frac{1}{q-1}\\
   &\times \sum_{\substack{u_i,u_i^\prime\in  {\mathbb{F}_q}\\
        u_i\ne u_i^\prime}}\sum_{y_1^N,u_1^{i-1}}\sqrt{
      P_N^{(i)}(u_i,y_1^N,u_1^{i-1})
      P_N^{(i)}(u_i^\prime,y_1^N,u_1^{i-1})}\\
&=\frac{1}{q-1}\sum_{y_1^N,u_1^{i-1}}P(y_1^N,u_1^{i-1}) \\
&\times    \sum_{\substack{u_i,u_i^\prime\in  {\mathbb{F}_q}\\
        u_i\ne u_i^\prime}}\sqrt{ P_N^{(i)}(u_i|y_1^N,u_1^{i-1})
      P_N^{(i)}(u_i^\prime|y_1^N,u_1^{i-1})}, 
    \end{aligned}
  \label{eq:1e}
\end{equation}
where the superscript $(i)$ in $P_N^{(i)}$ has been introduced to
establish a parallelism between $P_N^{(i)}$ and the coordinate channel
$W_N^{(i)}$ in \cite{arikan09}.

The parameters $Z(U_i| Y_1^N,U_1^{i-1})$ and
$ H(U_i| Y_1^N,U_1^{i-1})$, $i=1,\cdots N$, polarize simultaneously as
it is shown in \cite{arikan10} for the binary case, in
\cite{sasoglu2012polar} for a p-ary source, and in \cite{mori14} for a
q-ary source. An upper bound to the rate of polarization was obtained
in \cite{korada10} for binary channels with a polarizing
transformation defined by a matrix $G$. The bound was defined with the
help of the partial distances defined in the same paper. In
\cite{sasoglu2012polar} the bound to the rate of polarization was
extended for sources with prime alphabets and in \cite{mori14} for
sources with alphabet $\mathbb{F}_q$. The basic polar transformation
\cite{arikan10} and the extended transformation in \eqref{eq:1} have
associated matrices with the same partial distances. Therefore they
have the same bound for the rate of polarization. From
\cite{sasoglu2012polar} the rate of polarization for any $\beta$,
$0<\beta<1/2$ is,
\begin{equation}
  \label{eq:19}
  \lim_{N\to \infty}\frac{1}{N}|\{i:Z(U_i|Y_1^N,U_1^{i-1})\le 2^{-N^\beta}\}|=1-H(X|Y)
\end{equation}

%%%OJO QUITAR
%\vspace{1em}

\section{Decompression Via Recursive Evaluation of the LR}
\label{sec:recurs-eval-lr}

In this Section we present a  novel recursive expression for the LR
vector of the encoded symbols given the side information that
allows an effective implementation of the decoder. We also express the
Bhattacharyya parameter of $U_i$ given $(Y_1^N,U_1^{i-1})$ as a
function of the LR. With the Bhattacharyya parameters the construction of the q-ary
polar codes is done in a effective way using Monte Carlo. The LR
vector is defined in a similar way as in \cite{wymeersch04}. For its construction we choose
the additive identity of $\mathbb{F}_q$, the element ``0'',  as a reference and perform the ratio of the
probabilities of the remaining elements to the probability of the
reference one. As a consequence, the length of the likelihood ratio vector is
$q-1$. The components of the LR vectors
$\mathbf{L}_N^{(i)}(\cdot|y_1^N,u_1^{i-1})$, $i=1\cdots N$, are,
\begin{equation}
  \label{eq:10}
  L_N^{(i)}(u|y_1^N,u_1^{i-1})=\frac{P_N^{(i)}(u|y_1^N,u_1^{i-1})}%
{P_N^{(i)}(0|y_1^N,u_1^{i-1})};\ u\in \mathbb{F}_q, \ u\ne 0.
\end{equation}
Each component of $\mathbf{L}_N^{(i)}(\cdot|y_1^N,u_1^{i-1})$ is associated
with a non null element of $\mathbb{F}_q$.
%{\frac{N}{2}}

{\em Claim 1}: The LR vectors can be obtained recursively,
\begin{equation}
  \label{eq:11}
  %\begin{gathered}
\begin{split}
     L_{N}^{(2i-1)}&(u|y_1^{N},u_1^{2i-2})=\\
&\mkern-70mu \frac{\sum_{u_{2i}} %
              L_{\frac{N}{2}}^{(i)}(u-\alpha u_{2i}|y_{1}^{{\frac{N}{2}}},u_{1,o}^{2i-2} -\alpha u_{1,e}^{2i-2}) %
              L_{\frac{N}{2}}^{(i)}(u_{2i}|y_{{\frac{N}{2}}+1}^{N},u_{1,e}^{2i-2})}%
            { \sum_{u_{2i}} L_{\frac{N}{2}}^{(i)}(-\alpha u_{2i}|y_{1}^{{\frac{N}{2}}},u_{1,o}^{2i-2}-\alpha u_{1,e}^{2i-2}) %
              L_{\frac{N}{2}}^{(i)}(u_{2i}|y_{{\frac{N}{2}}+1}^{N},u_{1,e}^{2i-2})},\\
%\end{split}\\
%\begin{split}
     L_{N}^{(2i)}&(u|y_1^{N},u_1^{2i-2},u_{2i-1})=\\
&\mkern-54mu \frac{%
              L_{\frac{N}{2}}^{(i)}(u_{2i-1}-\alpha u|y_{1}^{{\frac{N}{2}}},u_{1,o}^{2i-2}-\alpha u_{1,e}^{2i-2})}%
            {L_{\frac{N}{2}}^{(i)}(u_{2i-1}|y_{1}^{{\frac{N}{2}}},u_{1,o}^{2i-2}-\alpha u_{1,e}^{2i-2})}%
      L_{\frac{N}{2}}^{(i)}(u|y_{{\frac{N}{2}}+1}^{N},u_{1,e}^{2i-2}),\\%
\end{split}
%%                L_N^{(i)}(u_{2i}|y_{N+1}^{2N},u_{1,e}^{2i-2})
%%L_N^{(i)}(u_{2i}|y_{N+1}^{2N},u_{1,e}^{2i-2})
  %\end{gathered}
\end{equation}
for $u\in \mathbb{F}_q, u\ne 0$, where $i=1\cdots \frac{N}{2}$, and a negative index of a conditioning variable
represents that there is not dependence on this variable. The recursion starts with $N=1$.

{\em Proof:} See Appendix A.

From the LR vector the
detection is performed in two steps: a) the component of
$\mathbf{L}_N^{(i)}(\cdot|y_1^N,u_1^{i-1})$ with maximal value is
obtained; b) if this component is larger that 1 the associated symbol is output. Otherwise the symbol ``0'' is output.

%
%%%%%%%%%%%%%%%%%%%%%%%%%
%%%%%%%%%%%%%%%%%%%%%%%%%%% 

\subsection{Bhattacharyya parameter}
\label{sec:cod-constr}
Binary polar codes can be constructed from the LR in the decoding
process \cite{arikan09}. The similarities between the LRs in
\cite{arikan09} and the recursive LRs in \eqref{eq:11} suggest that
the Bhattacharyya parameters can be expressed as a function of the
LRs. To this end we multiply and divide
the second term of \eqref{eq:1e} by $P_N^{(i)}(\beta|y_1^N,u_1^{i-1})$, where
$\beta \in \mathbb{F}_q$,
\begin{equation}
  \begin{aligned}
    Z&(U_i| Y_1^N,U_1^{i-1})=\frac{1}{q-1}\sum_{y_1^N,u_1^{i-1}}P(y_1^N,u_1^{i-1}) P_N^{(i)}(\beta|y_1^N,u_1^{i-1})\\
    &\qquad \times\sum_{\substack{u_i,u_i^\prime\in  {\mathbb{F}_q}\\
        u_i\ne u_i^\prime}}
    \sqrt{ \frac{%
      P_N^{(i)}(u_i|y_1^N,u_1^{i-1})
      P_N^{(i)}(u_i^\prime|y_1^N,u_1^{i-1})}%
    { P_N^{(i)}(\beta|y_1^N,u_1^{i-1})
      P_N^{(i)}(\beta|y_1^N,u_1^{i-1})}}
\end{aligned}
\label{eq:15a}
\end{equation}
After multiplying and dividing the second term of \eqref{eq:15a} by
$P_N^{(i)}(0|y_1^N,u_1^{i-1})$ we have,
\begin{equation}
\begin{aligned}
    Z&(U_i| Y_1^N,U_1^{i-1})\\
&=\frac{1}{q-1}\sum_{y_1^N,u_1^{i-1}}P(y_1^N,u_1^{i-1})
P_N^{(i)}(\beta|y_1^N,u_1^{i-1})
\frac{1}{        L_N^{(i)}(\beta|y_1^N,u_1^{i-1})}\\
    &\qquad \times%
\underbrace{%
\sum_{\substack{u_i,u_i^\prime\in  {\mathbb{F}_q}\\
        u_i\ne u_i^\prime}}%
%    \underbrace{%
      \sqrt{ %
        L_N^{(i)}(u_i|y_1^N,u_1^{i-1})
        L_N^{(i)}(u_i^\prime|y_1^N,u_1^{i-1})}}_{%
        (p-1)\Delta(y_1^N,u_1^{i-1})}\\
        &=\sum_{y_1^N,u_1^{i-1}}P(y_1^N,u_1^{i-1},u_i=\beta)
        \frac{\Delta(y_1^N,u_1^{i-1})}
{ L_N^{(i)}(\beta|y_1^N,u_1^{i-1})}\\
        &=\sum_{y_1^N,u_1^{N},u_i=\beta}P(y_1^N,u_1^{N})
        \frac{\Delta(y_1^N,u_1^{i-1})}
{ L_N^{(i)}(\beta|y_1^N,u_1^{i-1})},
    \end{aligned}
  \label{eq:15b}
\end{equation}
where $\Delta(y_1^N,u_1^{i-1})$ is defined as indicated in
\eqref{eq:15b}.

From \eqref{eq:15b} the Bhattacharyya
parameters are calculated using Monte Carlo from the LR vector used for
decoding  as in polar coding for the binary channel \cite{arikan09}. %
In \eqref{eq:15b} $Z(U_i| Y_1^N,U_1^{i-1})$ has been obtained for a fixed value
$u_i=\beta$. In the numerical simulations we estimate $Z(U_i|
Y_1^N,U_1^{i-1})$ by averaging the results obtained
over all possible values of $u_i$.
In \cite{arikan09} the construction of binary polar codes is based on
the Bhattacharyya parameters. From  \eqref{eq:15b} the construction
of q-ary polar codes  can be carried out in an effective way.

In the next Section we discuss the construction and implementation of
q-ary polar codes for sources and channels.

\section{Coding with q-ary Polar Codes}
\label{sec:source-coding}

The polar transformation presented in \eqref{eq:1} for two variables,
and extended in Fig. \ref{fig:1} for $N$ variables, can be used for source
coding with side information, or for channel coding
\cite{arikan10}. Though the problem is very similar in both cases,
there are, however, subtle differences.

\subsection{Source Coding Implementation and Code Construction}
\label{sec:source-coding-1}

The information to be compressed is
$X_1^N$ and  $Y_1^N$ is the side information available for
decoding. Let $\mathcal{A}$ be the set of indices with low
Bhattacharyya parameters,
\begin{equation}
  \label{eq:14}
 \mathcal{A}=\{i : Z(U_i| Y_1^N,U_1^{i-1})<\delta\},
\end{equation}
with $| \mathcal{A}^c|=\lceil R_s N\rceil$, $R_s>H(X|Y)$. $R_s$ is the code
rate for source compression and $\lceil \cdot \rceil$ is the ceil
function of its argument. The set $\mathcal{A}$ is called information set in \cite{arikan09}
and the set $\mathcal{A}^c$ is called high entropy index
set \cite{arikan10}, as it gathers the variables with high conditional entropy. The
variables $u_\mathcal{A}$ have low conditional entropy. They can be
recovered using SC decoding with \eqref{eq:11}. However, the
variables $u_{\mathcal{A}^c}$, the frozen variables, have high entropy
and are not recoverable with SC decoding. Therefore,
$u_{\mathcal{A}^c}$ is the compressed pattern of $X_1^N$.

The detection is done in two steps: 1) recovering of $U_1^N$ from
$u_{\mathcal{A}^c}$ and $Y_1^N$, 2) obtaining $X_1^N$ from $U_1^N$.

\subsubsection{Recovering $U_1^N$}
\label{sec:recovering-u_1n}

We have to detect the components of $U_1^N$ that are in the information
set $\mathcal{A}$. The rest, $U_{\mathcal{A}^c}$, are available. As we already know $Y_1^N=y_1^N $, we can obtain $P(x_i|y_i)$,
$x_i\in \mathbb{F}_q$. Therefore, for each coordinate $y_i$ we
can calculate the LR vector
$\mathbf{L}_1^{(i)}(\cdot | y_i)$ with components
\begin{equation}
  \label{eq:17}
  L_1^{(i)}(u|y_i)=\frac{P(u|y_i)}{P(0|y_i)}; \ u\in \mathbb{F}_q, \
  u\ne 0.
\end{equation}
After the initial step \eqref{eq:17} the recursion \eqref{eq:11} is
repeated $\log_2 N$ times. The LR vectors with indices in $\mathcal{A}$ are
detected as indicated in Section III.

\subsubsection{Obtaining $X_1^N$}
\label{sec:obtaining-x_1n}
The application of the inverse transform $G_N^{-1}$ to $U_1^n$ gives
$X_1^N$, i.e.,
\begin{equation}
  \label{eq:18}
  X_1^N=U_1^{N} G_N^{-1}.
\end{equation}
For a binary polar code $G_N=G_N^{-1}$. However, this is not true for a
general q-ary polar code.

The information set $\mathcal{A}$ allows us to upper bound the error
probability. If the detected pattern is $\hat{X}_1^N$, the error probability is,
\begin{equation}
  \label{eq:21}
  P_e=Pr(\hat{X}_1^N\ne X_1^N). 
\end{equation}
From \cite{arikan09} and \cite{sasoglu2012polar} we can show that the error probability is bounded by the
Bhattacharyya parameters associated to the information set,
\begin{equation}
  \label{eq:20}
  P_e\le (q-1)\sum_{i\in \mathcal{A}}Z(U_i|Y_1^N,U_1^{i-1}).
\end{equation}
Thus, the sum $\sum_{i\in \mathcal{A}}Z(U_i|Y_1^N,U_1^{i-1})$ is an important
parameter for the code construction.

The polar source coding reaches the theoretical limit, $R_s\to H(X|Y)$
as $N\to \infty$, as it can be seen from \eqref{eq:19} and \eqref{eq:14}.

\subsection{Channel Coding}

\begin{figure}[htb]
 \centering
 \includegraphics[width=8.5cm]{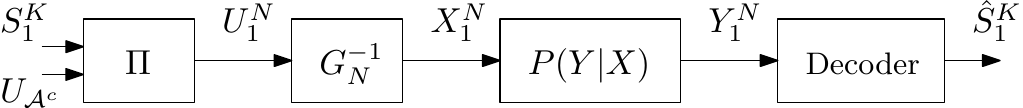}
 \caption{Channel coding.}
\label{fig:2}
\end{figure}

%decir que en In \cite{sasoglu se demuestra que la suma de los Z sobr
%cal A es <Pe.
The problem of channel coding can be view as an instance of source
coding where the side information $Y_1^N$ is the channel output. The
input to the channel $X_1^N$ can be understood as the information to
be compressed in the source coding problem. %
Therefore, the information to be transmitted to the channel $S_1^K$
must have a similar role as the symbols that can be recovered with SC
decoding in the source coding case, $U_\mathcal{A}=S_1^K$. In the
source coding problem the high entropy symbols $u_{\mathcal{A}^c}$ are
available to both the encoder and the decoder. In channel coding this
not possible, in general. One way to overcome this is to generate the
frozen symbols with synchronized pseudorandom generators on the
transmitter and receiver sides of the channel. $X_1^N$ is obtained
through an inverse polar transform $G_N^{-1}$ of the combined
vector formed from the message and frozen symbols.  Fig. \ref{fig:2}
depicts the scheme of the channel polar coding system, where the block
$\Pi$ is a permutation that produces $U_1^N$ from the message and the
frozen symbols.

%%%OJO AQUI ESTABA LA FIGURA 2

%%%%ojo pensarlo
The code rate for channel coding is $R_c=|\mathcal{A}|/N$ and, if the
message symbols are independent and identically distributed (iid)
then, as $N\to \infty$ \cite{sasoglu2012polar}
\begin{equation}
  \label{eq:23}
  R_c\to 1-H(X|Y),
\end{equation}
where the log in the entropy is base q. The q-ary polar code can reach the symmetric
capacity, but not the real capacity; however, by increasing the source
alphabet it is possible to apply some procedures to reach it \cite{sasoglu2012polar}.

%%%%%%%%%%%INTRODUCIRLO AQUI

A relation between channel degradation and the
Bhattacharyya parameters of binary channels was presented in \cite{korada09}. The
definition of channel degradation \cite{korada09} is.

{\em Definition 2}: Let $P^{(1)}(y_1|x): \mathbb{F}_q\rightarrow
  {\cal{Y}}_1$ and $P^{(2)}(y_2|x):  \mathbb{F}_q\rightarrow
  {\cal{Y}}_2$ two q-ary discrete memoryless channels (DMC). $P^{(1)}$ is
  degraded with respected to $P^{(2)}$, $P^{(1)}\preceq P^{(2)}$, if
  there is a channel $P(y_1|y_2): {\cal{Y}}_2 \rightarrow
  {\cal{Y}}_1$ such that,
  \begin{equation}
    \label{eq:deg_chan}
    P^{(1)}(y_1|x)=\sum_{y_2\in \cal{Y_2}} P^{(2)}(y_2|x) P(y_1|y_2)
  \end{equation}
For channel coding we consider the Bhattacharyya parameter
$Z(U_i|U_1^{i-1},Y_1^N)$, where $U_1^N$ is the vector of symbols at
the input of the q-ary encoder $G_N^{-1}$, as indicated in
Fig. \ref{fig:2}. In the following claim we extend to the q-ary case the relation in \cite{korada09}
between  channel degradation and 
Bhattacharyya parameters of binary channels.

{\em Claim 2}: Given two channels $P^{(1)}$ and $P^{(2)}$,  with $P^{(1)}\preceq P^{(2)}$, then
$Z(U_i|U_1^{i-1},{Y_1}_1^N)\ge Z(U_i|U_1^{i-1},{Y_2}_1^N)$.

Proof: See Appendix B.

The relation between channel degradation and the Bhattacharrya
parameters of the coordinate channels \cite{arikan09} has been used to apply binary
polar codes to relay channels \cite{andersson10,bravo13rel}.  In
\cite{bravo13ray} a rate adaptation technique is applied to Rayleigh
channels with binary polar coding. With the help of Claim 2 it is straightforward to
extend the commented binary applications to the q-ary case.

For polar channel coding on the AWGN we need to
establish a map between $\mathbb{F}_q$ and a set of signals. In this paper a
signal is a complex number that represents a complex function of time. Therefore, for the
constellation $C=\{t_j : j=0,\cdots,q-1; t_j\in \mathbb{C}\}$ we have
the map $x_j\mapsto t_j$, $x_j\in \mathbb{F}_q$,
$t_j\in \mathbb{C}$, $j=0,\cdots q-1$.

Let the noise of the AWGN channel be the random variable $Z$, and
$Z_r=\Re{(Z)}$, $Z_i=\Im{(Z)}$, with $\Re(\cdot)$ and $\Im(\cdot)$ the
real and imaginary parts, respectively, of the argument, and let the
mean square of the noise be $E[|Z|^2]=2\sigma^2$. The probability
density function (pdf) of $Z$ is \cite{lapidoth09}

\begin{equation}
  \label{eq:pdfz}
  f_Z(z)=f_{Z_r,Z_i}(z_r,z_i)=\frac{1}{2\pi \sigma^2} e^{-\frac{1}{2\sigma^2}(z_r^2+z_i^2)}.
\end{equation}

The signals at the output of the channel are, $Y_i=t(x_i)+Z_i$ $i=1\cdots
N$. As the symbols
$x_i$, $x_i \in \mathbb{F}_q$, are equally likely, then the components
of the LR vector for the
first step of the iteration $ \mathbf{L}_1^{(1)}(\cdot|y_i)$,
$i=1,\cdots,N$ are

\begin{equation}
  \label{eq:24}
  L_1^{(1)}(x|y_i)=\frac{f_Z\left( y_i-t(x)\right)}{f_Z\left(
      y_i-t(0)\right)}; x\in \mathbb{F}_q, x\ne 0
\end{equation}

The log LR (LLR) of \eqref{eq:24} is,
\begin{equation}
  \label{eq:25}
\begin{split}
  l_1^{(1)}&(x|y_i)\\
&=\frac{\Re(t(x))-\Re(t(0))}{\sigma^2}\left(\Re(y_i)-\frac{1}{2}\big(\Re(t(x))+\Re(t(0))\big)\right)\\
&+\quad \frac{\Im(t(x))-\Im(t(0))}{\sigma^2}\left(\Im(y_i)-\frac{1}{2}\big(\Im(t(x))+\Im(t(0))\big)\right),
\end{split}
\end{equation}
where $x\in \mathbb{F}_q, x\ne 0$.

 We consider two kind of
constellations, rectangular shaped and circular shaped.

\subsubsection{Rectangular Shaped Constellations}
\label{sec:rect-shap-const}

\begin{figure}[h]
\centering
\begin{tikzpicture}
%\node (n0) at (0,0) {\includegraphics[width=4cm]{fgcir_49}};
%\node (n1) at (5,0) {\includegraphics[width=4cm]{fgcir_64}};
\node (n0) at (0,0) {\includegraphics[width=3.5cm]{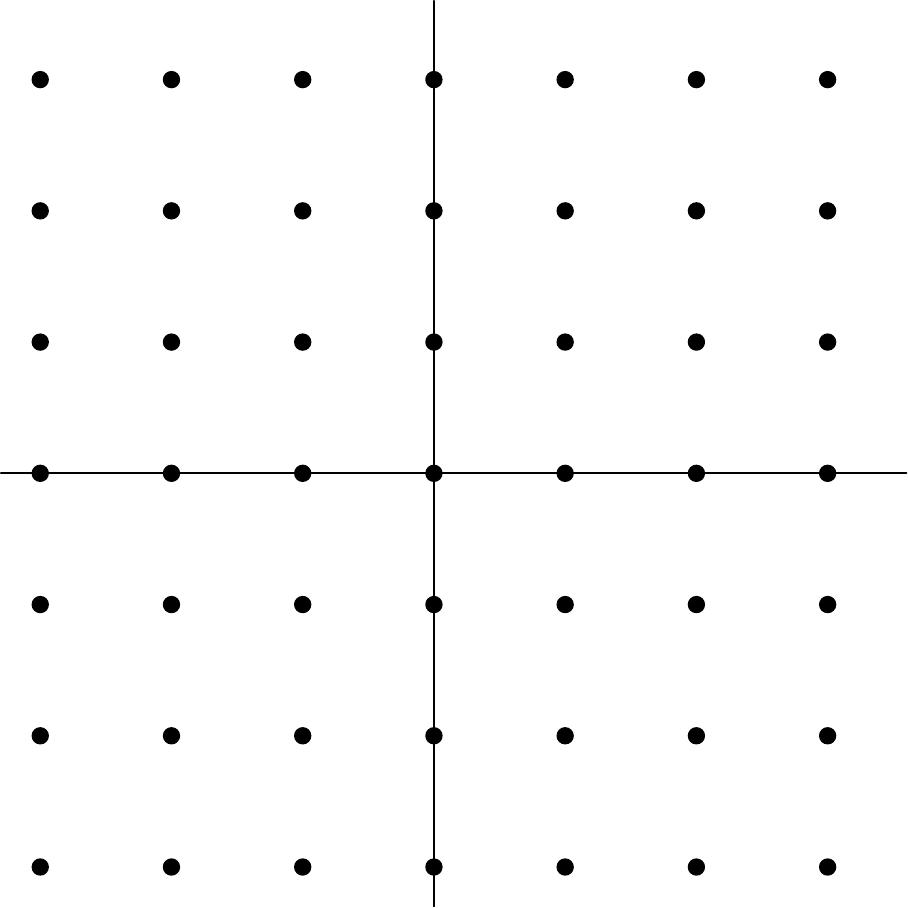}};
\node (n1) at (4,0) {\includegraphics[width=3.5cm]{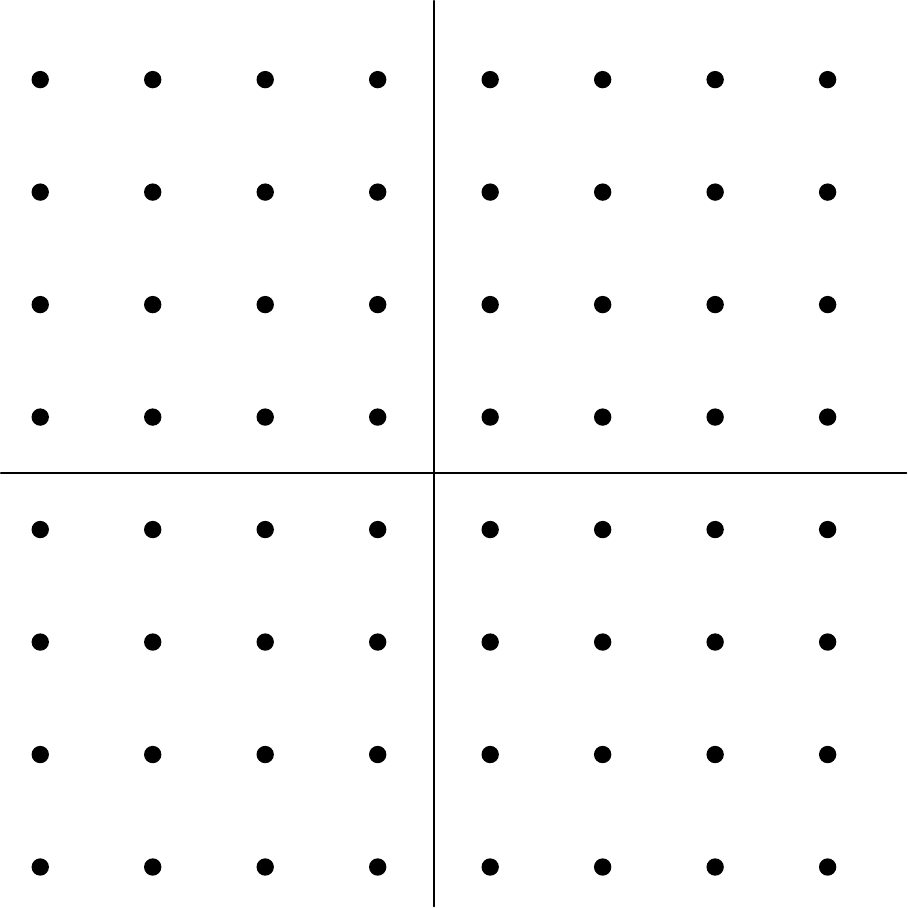}};
\node (lab0)[below=1ex of n0]{a) 49-QAM};
\node (lab1)[below=1ex of n1]{b) 64-QAM};
\end{tikzpicture}
  \caption{Rectangular QAM constellations.}
\label{fig:2a}
\end{figure}
%\begin{figure}[h]
% \centering
% \includegraphics{fgcir_49}
% \caption{Rectangular QAM constellation}
%\label{fig:2a}
%\end{figure}

They are formed by spreading points equally spaced along the real and imaginary axis. We form a PAM constellation along each axis. If the number of points per axis is prime, the PAM constellation  is,
$C_{pPAM}=\{i-\lfloor p/2 \rfloor:i=0,\cdots,p-1\}$, where $\lfloor
\cdot \rfloor$ is the floor function. If the size of the PAM
constellation is $2^m$, the constellation is,
$C_{2^mPAM}=\{2i-(2^m+1):i=1,\cdots,2^m\}$. It is possible to work
with two different alternatives with rectangular shaped
constellations. One alternative is to use two independent PAM
constellations, each with its own polar code, one in the real axis and
the other in the imaginary axis, in order to obtain the
rectangular quadrature amplitude modulation (QAM) constellation. The
main advantage of this alternative is its efficiency. The other
alternative is to use all the points of the rectangular
QAM constellation with one q-ary polar code. This alternative has
better performance, but is less efficient than the previous one. 

It is difficult to work with prime PAM constellations with size larger than 13 symbols, as it
is shown in Section V. However, PAM constellation with size $2^m$,
$m\in \mathbb{N}$, do not have the same problems and in Section V we work with two
32-PAM constellations to construct a 1024-QAM constellation. In
Fig. \ref{fig:2a},  a) and b), we show the constellations 49-QAM and
64-QAM, respectively. Both of them have been used in the simulations.

\subsubsection{Circular Shaped Constellation}
\label{sec:circ-shap-const}

\begin{figure}[ht]
  \centering
  \includegraphics[width=8.5cm]{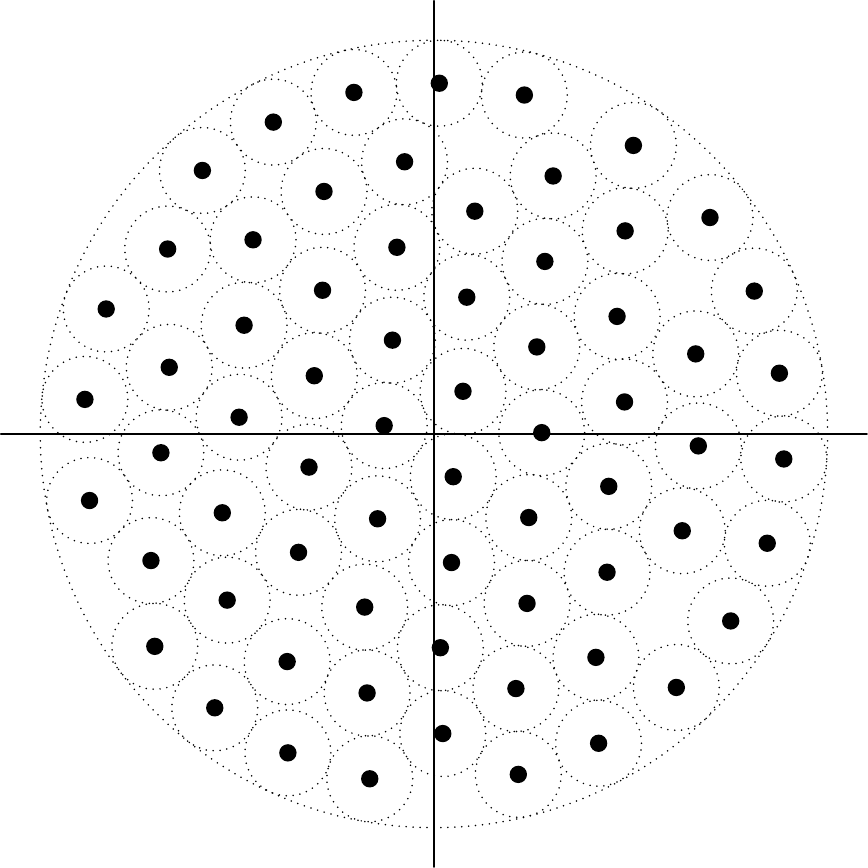}
  \caption{Circular 67-QAM constellation.}
\label{fig:2b}
\end{figure}

The problem of maximizing the minimum pairwise distance between points of a
constellation of a fixed energy, whose points are equally likely, is equivalent to the
problem of optimal packing of equal circles in a circle,
where optimal means non-overlapping circles of maximum radius
\cite{graham1998dense}. In \cite{spetche2009} a data base is maintained
of the best
packings of circles in a circle up to a size of $2600$ circles. By
choosing a packing with a number of circles equal to the order of a field we can define a
constellation formed by the center of the circles. As the points of
the constellation are spread in a circle we refer to it as circular
QAM constellation. In Fig. \ref{fig:2b} we show the circular 67-QAM
constellation, that it is one of the constellations used in this
paper. The interest of the 67-QAM constellation is more theoretical
than practical, as a small rate loss is expected from mapping
binary information to symbols. However, the results show that this
constellation is close to the theoretical limit, compensating the
commented rate loss. Additionally, efficiency is gained from
working in a prime field, with addition module a prime,  as compared to
that in a non-prime field, where the addition is more complex.

\subsection{Encoder and Decoder Implementation}

The equations \eqref{eq:1} and \eqref{eq:11}, that describe the q-ary
polar encoder and decoder respectively, have a
similar structure to the equations defining the binary polar encoder and
decoder. Therefore the implementation of the q-ary polar encoder and
decoder follows similar principles as in the binary case.

The decoder calculates the LRs \eqref{eq:11} 
in order to carry out SC decoding. Eq. \eqref{eq:11} is recursive and,
in order to show the dependency of the current LRs on the previous ones,
we express them as,
%{\frac{N}{2}}
\begin{equation}
  \label{eq:recurlr}
  \begin{aligned}
     \mathbf{L}_{N}^{(2i-1)}&=f\big(\mathbf{L}_{\frac{N}{2}}^{(i)}(u_{1,o}^{2i-2} -\alpha u_{1,e}^{2i-2}),\mathbf{L}_{\frac{N}{2}}^{(i)}(u_{1,e}^{2i-2})\big)\\
&=f\big(\mathbf{L}_{{\frac{N}{2}},oe}^{(i)},\mathbf{L}_{{\frac{N}{2}},e}^{(i)}\big)\\
     \mathbf{L}_{N}^{(2i)}&=g\big(\mathbf{L}_{\frac{N}{2}}^{(i)}(u_{1,o}^{2i-2} -\alpha u_{1,e}^{2i-2}),\mathbf{L}_{\frac{N}{2}}^{(i)}(u_{1,e}^{2i-2},u_{2i-1})\big)\\
&=g\big(\mathbf{L}_{{\frac{N}{2}},oe}^{(i)},\mathbf{L}_{{\frac{N}{2}},e}^{(i)},u_{2i-1}\big),
\end{aligned}
\end{equation}
where $\mathbf{L}_{{\frac{N}{2}},oe}^{(i)}$, $\mathbf{L}_{{\frac{N}{2}},e}^{(i)}$, $f()$ and $g()$ have been introduced to highlight the functional dependency of the LRs in the actual recursion on the LRs  of the previous one.

\begin{figure}[ht]
  \centering
  \includegraphics{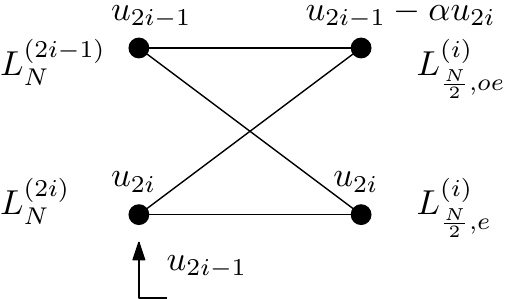}
  \caption{Butterfly for decoding}
\label{fig:butterfly}
\end{figure}

The structure of \eqref{eq:recurlr} resembles that of the basic unit
for calculating the fast Fourier transform (FFT): the butterfly structure. In
Fig. \ref{fig:butterfly} we represent the butterfly diagram of
\eqref{eq:recurlr}. A
similar structure is also used for decoding binary polar codes
\cite{arikan09}. There are two kind of calculations in
\eqref{eq:recurlr} and in Fig. \ref{fig:butterfly}: from right to
left, forward direction, the LRs of the actual recursion are
calculated with \eqref{eq:recurlr} as a function of the LRs from the
previous iteration. From left to right, backward direction, the extended
transformation \eqref{eq:1} is inverted. The even LR vector, $\mathbf{L}_{2N}^{(2i)}$, is a
function of the symbol $u_{2i-1}$ and this fact is indicated in Fig. \ref{fig:butterfly} by the
arrow in lower left side of the butterfly. The values calculated in the forward direction, two vectors of real numbers, and backward direction, two elements of $\mathbb{F}_q$, must be stored for their use by other butterflies.

\begin{figure}[ht]
  \centering
  \includegraphics[width=6cm]{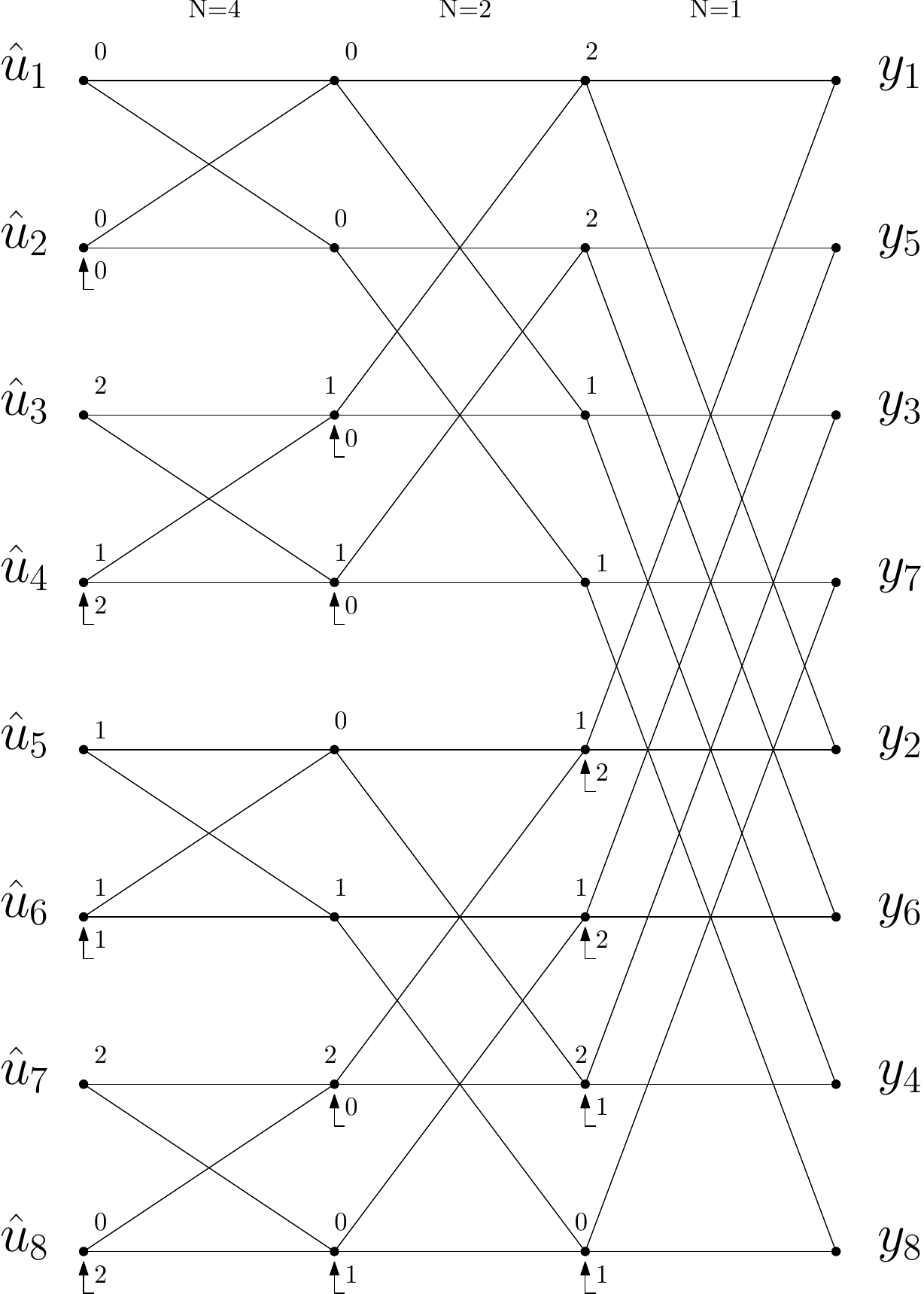}
  \caption{Butterfly diagram}
\label{fig:butt_dia}
\end{figure}

We use an example to describe the algorithm implemented in the
decoder. The codeword length is N=8 and the symbol alphabet is
$\mathbb{F}_3$, with elements \{0,1,2\} and modular arithmetic modulus
3. In Fig. \ref{fig:butt_dia} we present the butterfly diagram used
for decoding. On the right of the diagram are the signals $y_i$,
$i=1\cdots 8$, at the input of the decoder. They are applied to the
decoder after a bit reversal permutation. The LR vectors of these
signals are calculated with \eqref{eq:17}, or with \eqref{eq:24}, and
they are the input to the rightmost butterflies. The detected symbols
are on the left out of the diagram. Decoding starts with
the first symbol, $\hat{u}_1$, continues with the second when the
first has been detected, and so on. The butterfly associated to the
first symbol checks if it has LR vectors at its input. If it is the
case the butterfly obtains the LR and performs the
detection. Otherwise it transfers control to its neighboring
butterflies on the right, first the upper one and, when it receives
the LR vector, to the lower one. This process of transferring control
is repeated from left to right until a butterfly is able to calculate
the LR vector. In this case the butterfly gives the LR vector to its
left neighbor and transfers control to it. When the upper leftmost
butterfly has both LR vectors available it first obtains $L_{8}^{(1)}$
and detects the symbol $\hat{u}_1$. With it the LR
$\mathbf{L}_{8}^{(2)}$ is calculated and $\hat{u}_2$ is detected. When
both symbols of a butterfly are detected, or they are available, it
does the inverse extended transformation \eqref{eq:1}. In
Fig. \ref{fig:butt_dia} we have indicated the detected symbols on the
left of the leftmost butterflies. On the right of the butterflies are
the symbols obtained after inverting the extended trasformation. These
symbols are needed by the inner butterflies. For instance, the
butterfly identified by its left upper corner as $\hat{u}_1$ and
recursion $N=2$, needs the symbol $\hat{u_1}-\hat{u}_2$ (0 in this
case) to obtain the even LRs. Therefore, as the symbols $\hat{u}_i$,
$i=1\cdots N$, are detected, the extended transformation is inverted
in the butterflies from left to right. As a result, the rightmost
butterflies give the vector $\hat{\mathbf{u}}^8 G_8^{-1}$. Therefore,
the butterfly diagram can be used for encoding too.

In this paper we have used a butterfly diagram for encoding. It avoids problems with the storage of the matrix $G_N$, or its inverse, and a better efficiency can be obtained as matrix multiplication is less efficient, in general.

\subsection{Complexity}
\label{sec:complexity}

From \eqref{eq:1} the complexity of the q-ary polar encoder is similar to the binary encoder: $N\log_2 N $. The butterfly diagram describes the complexity of the decoder. The
number of butterflies is $\frac{N}{2}\times\log_2 N$. From \eqref{eq:11}, in each butterfly the
odd LR needs $q$
operations per LR coordinate. The even LR only needs
one operation per coordinate. Therefore the complexity of decoding is
$\mathcal{O}(\frac{q^2}{2} N \log_2 N)$.

For decoding each butterfly stores the calculated LRs, with the
exception of the leftmost butterflies: they do not store the
obtained LRs, they
just detect. However,  the LRs at the input of the rightmost
butterflies must be stored. The butterflies must also store the
symbols obtained from the inverse transformation. If the memory for storing symbols is
considered negligible compared to that for storing real numbers, then
the decoder needs $\mathcal{O}((p-1)N\log_2 N$ cells for real numbers.

Many of the operations indicated in the butterfly diagram in Fig. \ref{fig:butt_dia} can be run
in parallel. For instance, in all the rightmost butterflies the odd
LRs $\mathbf{L}_{2}^{(1)}$ can be calculated in parallel. In their
neighboring butterflies, half of the odd LRs can be calculated in
parallel, and so on. And the odd LRs are the most complex, as
commented before. Likewise, the frozen symbols %%%ojo chequear que
                                %%%se ha definido
can help in the process of parallel computing. Obviously, the
scheduling of the operations among butterflies must be adapted for
parallelization.

\section{Numerical Results}
\label{sec:numerical-results}

We have performed several numerical experiments in order to corraborate
the theoretical results and to gain insight into the application of
q-ary polar codes. Two kind of experiments have been performed:
source coding simulations and channel coding for the AWGN channel.

\subsection{Source Coding with Side Information}
\label{sec:source-coding-with}

\begin{table}[h]
  \centering
  \begin{tabular}{c|ccccc|}
&\multicolumn{5}{|c|}{$x$} \\\cline{2-6}
&0&1&2&3&4\\\hline
$P_X(X)$&0.300 & 0.200 & 0.300 & 0.100 & 0.100\\\hline
\end{tabular}
 \caption{Probabilities of the source symbols.}
  \label{tab:1}
\end{table}

\begin{table}[h]
  \centering
  \begin{tabular}{c|c|ccccc|}
    \multicolumn{2}{c|}{ \multirow{2}{*}{$P(X|Y)$} } &  &  & y &  & \\\cline{3-7}
    \multicolumn{2}{c|}{} & 0 &  1 & 2 & 3 & 4\\\hline
    \multirow{5}{*}{$x$}&0&0.522 & 0.261 & 0.136 & 0.231 & 0.316\\
                                                  &1&0.261 & 0.348 & 0.092 & 0.154 & 0.105\\
                                                  &2&0.131 & 0.261 & 0.682 & 0.231 & 0.158\\
                                                  &3&0.043 & 0.087 & 0.045 & 0.307 & 0.105\\
                                                  &4&0.043 & 0.043 & 0.045 & 0.077 & 0.316\\\hline
  \end{tabular}
  \caption{Conditional probabilities of the source symbols.}
  \label{tab:2}
\end{table}

\begin{figure}[ht]
  \centering
  \includegraphics[width=9cm]{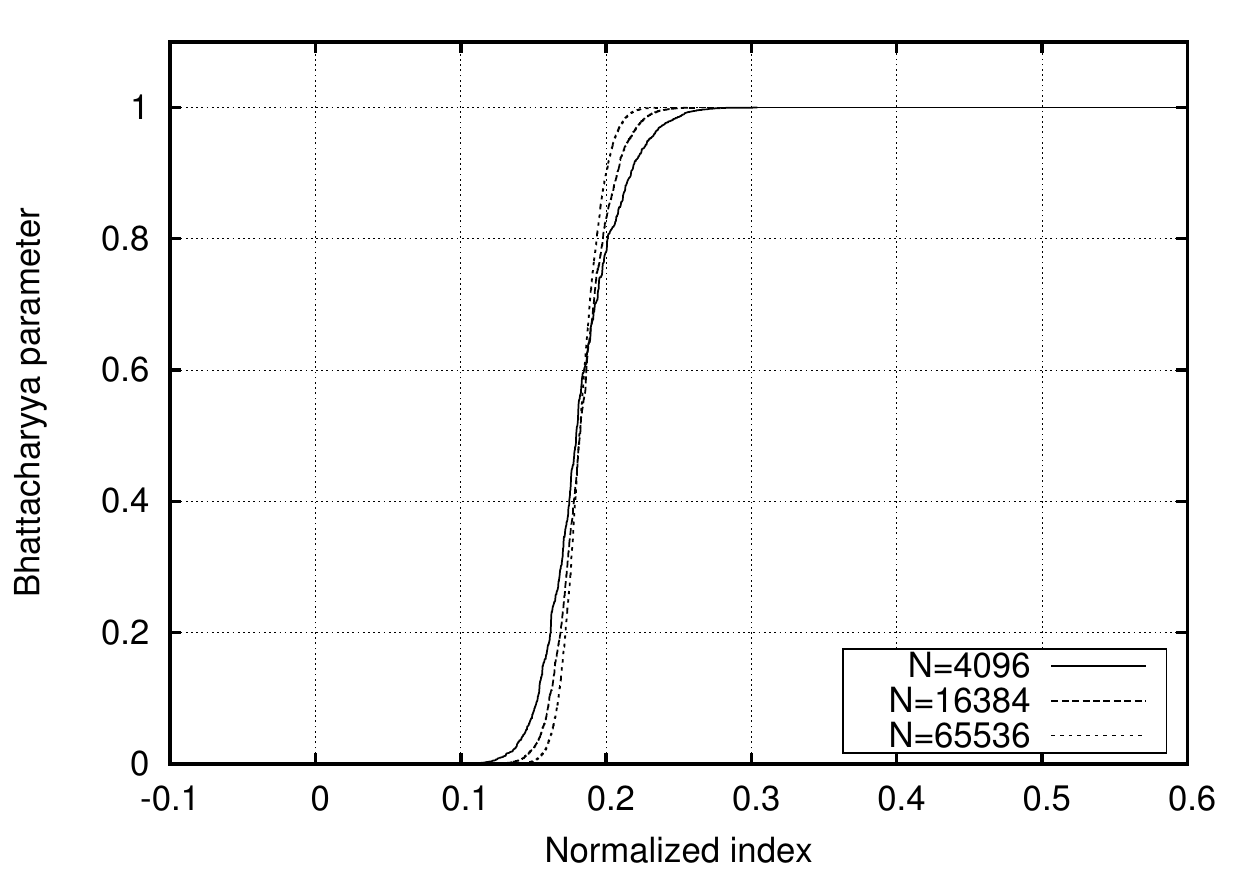}
  \caption{Bhattacharyya parameters of the source polar codes.}
\label{fig:3}
\end{figure}

Several simulations have been done with several sources in order to
know how close the results with finite codeword length are to the
theoretical limits. In tables \ref{tab:1} and \ref{tab:2} we gather the
probabilities associated to one of sources cosidered in the
experiments. The conditional entropy of the source is $H(X|Y)=1.90061$
bits. The source alphabet is $\mathbb{F}_5$. Three codeword lengths
have been chosen: $N=4096$, $N=16384$, and $N=65536$. The codes have been constructed using Monte Carlo, as
explained in Section III. The ordered Bhattacharyya parameters of the
codes are presented in
Fig. \ref{fig:3} as a function of the indices normalized by $N$. As
expected, the longer the code is, the steeper the slope of the curve.

For the code construction the set $\mathcal{A}$ was formed with
indices whose Bhattacharyya sum is upper bounded,
\begin{equation}
  \label{eq:sum2}
\sum_{i\in \mathcal{A}}Z(U_i|Y_1^N,U_1^{i-1})\le 10^{-4}.  
\end{equation}

\begin{figure}[ht]
  \centering
  \includegraphics[width=9cm]{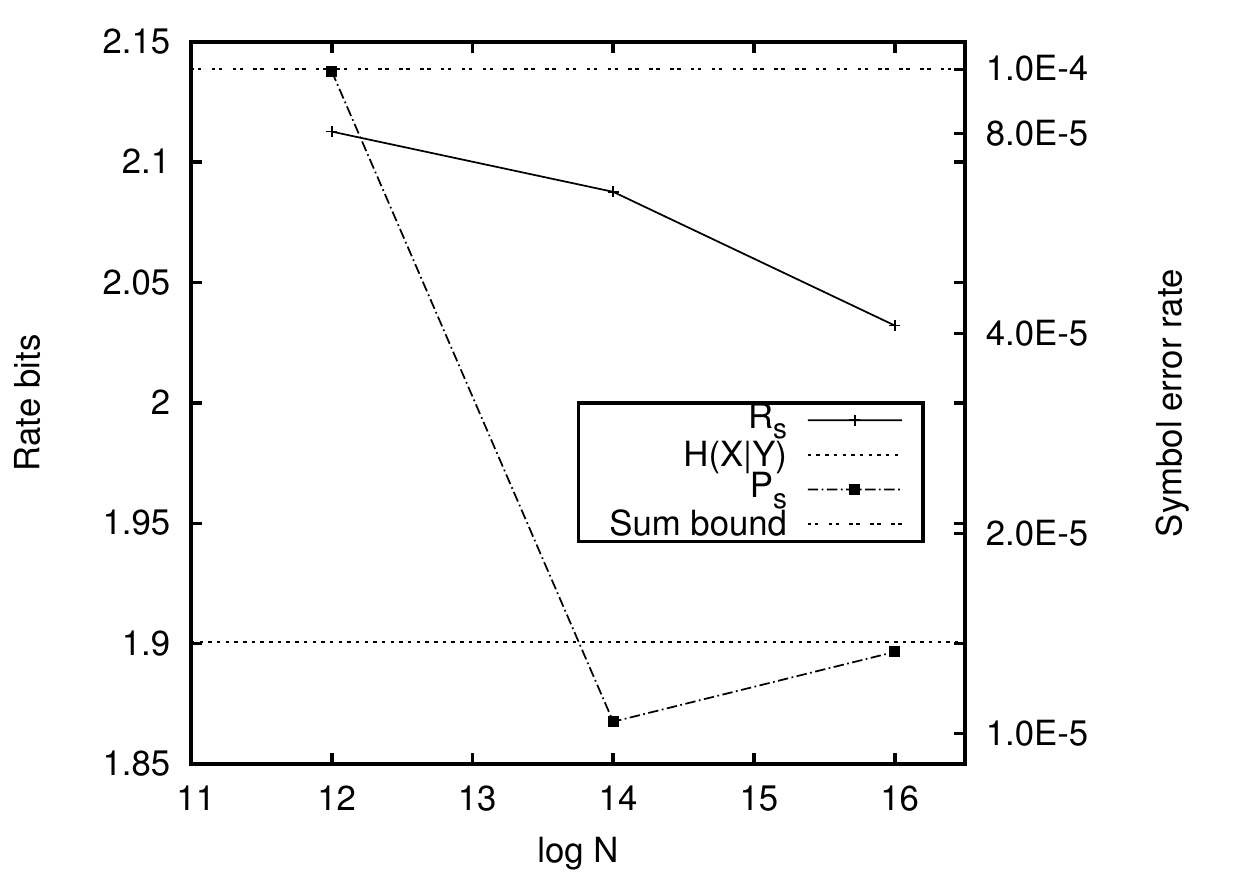}
  \caption{Source compression rate of three codes constructed with
    the sum bound criterion in \eqref{eq:sum2}. The x-axis shows the $\log_2$ of the
    codeword length. In the same figure it is shown the symbol
    error rate obtained with the codes.}
\label{fig:4}
\end{figure}

The source code rate $R_s$ is obtained from $|{A^c}|$. In Fig. \ref{fig:4} the
source rate of the three mentioned codes is plotted versus the
codeword length. In the same figure $H(X|Y)$ is represented. The
source rate gets closer to the entropy as the codeword length
increases. Also, in Fig. \ref{fig:4} we present the symbol
error rate, indicated by $P_s$ in the figure,  obtained by
counting symbol errors between the input to the encoder and the output from
the decoder. The upper bound 10${}^{-4}$ in \eqref{eq:sum2}, used for the code
construction, has also been represented in Fig. \ref{fig:4}. For the shortest code the simulation results are close to
the sum upper bound. However, for long codes the criterion for constructing
$\mathcal{A}$ using the sum bound is conservative.

\subsection{Channel Coding}
\label{sec:channel-coding}

\begin{figure}[ht]
  \centering
  \includegraphics[width=9cm]{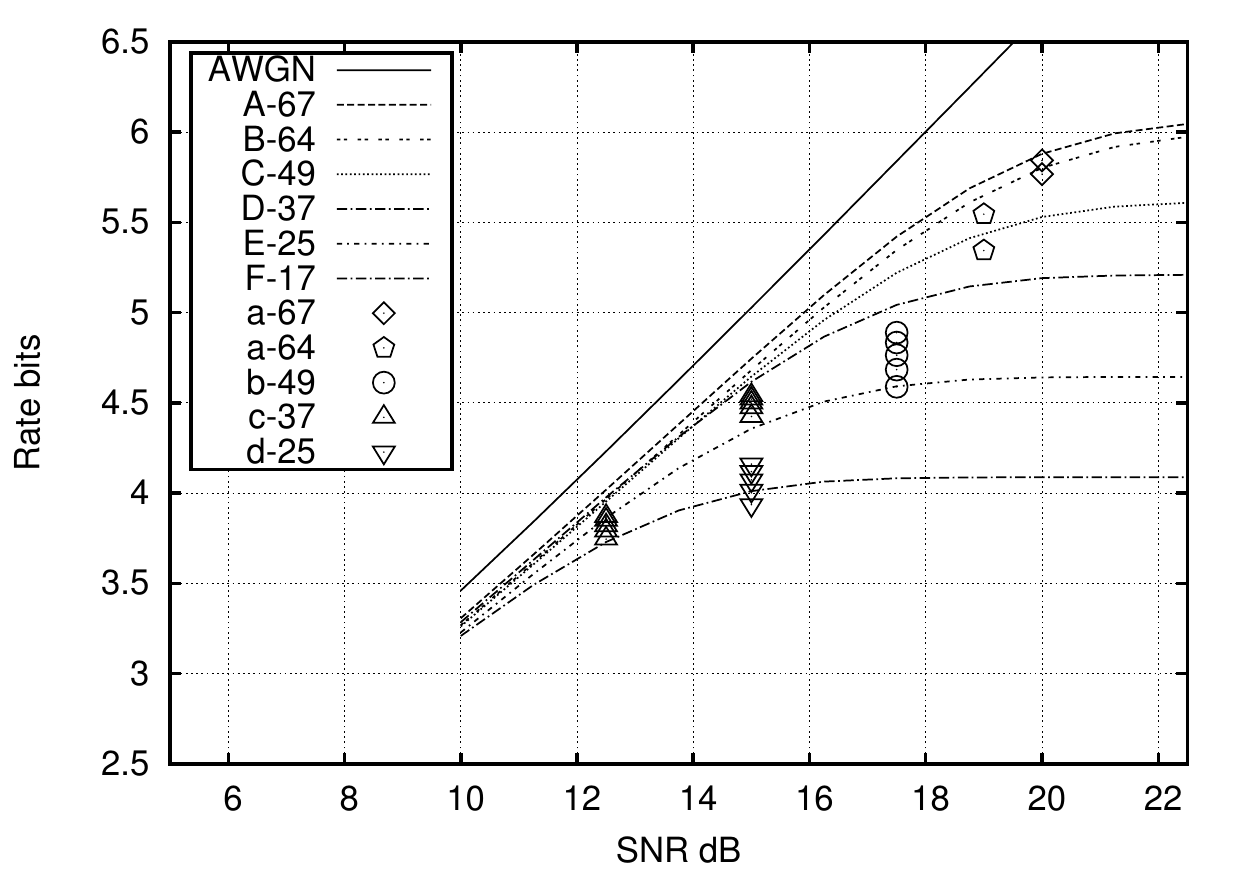}
  \caption{Theoretical rates and rates obtained in the
    simulations versus SNR in dB. The curves labelled with appercase
    letters are theoretical. The number following the letter is the
    constellation's size. For curves a-64 and a-67 the codeword
    lengths are in $\{2048,65536\}$. In the remaining simulations the
    codeword lengths are in $\{4096,8192,16384, 65536\}$.}
\label{fig:5}
\end{figure}

A set of simulations have been done in order to know the performance of q-ary
polar channel coding. Despite we have done simulations with
both discrete and continous channels, in this work we only present results
for the continous case: the AWGN channel. The signal noise ratio (SNR)
has been defined as $SNR=E_s/E[|Z|^2]$, where $E_s=E[|t(x)|^2]$.

The two constellations
presented in Section IV haven have been considered. The code
construction was done in a similar way as that for source coding,
using the LR for obtaining the Bhattacharyya parameters, as indicated
in Section III. Two channel coding alternatives have been used for rectangular QAM
constellations. In one of them each PAM constellation has been
considered independently for coding. In the other all the
constellation points are considered for coding. For the 
independent PAM constellations the LLR \eqref{eq:25} and
the noise variance are adapted for the case of real signals. Different
codeword lengths have been considered, from 2048 to 525288.

In the first set of experiments we obtained the information set
$\mathcal{A}$ for different constellations and codeword lengths. The
criterion for including an index in $\mathcal{A}$ was that its
Bhattacharyya parameter were less than a prefixed value,
instead of the sum Bhattacharyya parameters used for source
coding. The chosen value was $10^{-4}$. In Fig. \ref{fig:5} we show
the rates $R=R_c\log_2 q$, in bits, obtained with simulation
for different codeword lengths and SNRs. For
comparison, in the same figure we show the theoretical mutual
information curves associated to the constellations with equally likely constellation points. The curves have been
calculated numerically. The constellations are identified by their cardinality; if it is prime the constellation's shape is circular,
otherwise it is rectangular. In the case of rectangular QAM
constellations the coding has been performed on each PAM constellation
independently. The theoretical results are indicated with
uppercase letters. Two sets of codeword lengths have been considered
for the simulations $\{2048,65536\}$ for 64 and 67 signal points per
constellation and $\{4096,8192,16384, 65536\}$ for the remaining
constellations. Increasing the
codeword length for a fixed SNR and constellation, renders rates that
are closer to the theoretical results. Because of this we have not
labeled the codeword lengths: they are easily identified
by their position in the pile corresponding to the SNR and shape.

Three main conclusions can be extracted from the results in
Fig. \ref{fig:5}, a) the circular QAM shaped constellation gives
results that are close to the theoretical curves, even with short
codeword lengths, b) the rectangular QAM constellation with size
$2^m$, $m\in \mathbb{N}$, is close to the theoretical curves for short
codes, though the
separation from the theoretical curves is noticeable, c) the
rectangular QAM shaped constellation with size $p^2$, $p$ prime, is further from the theoretical results and the gap
increases with the constellation size.

\begin{figure}[ht]
  \centering
  \includegraphics[width=9cm]{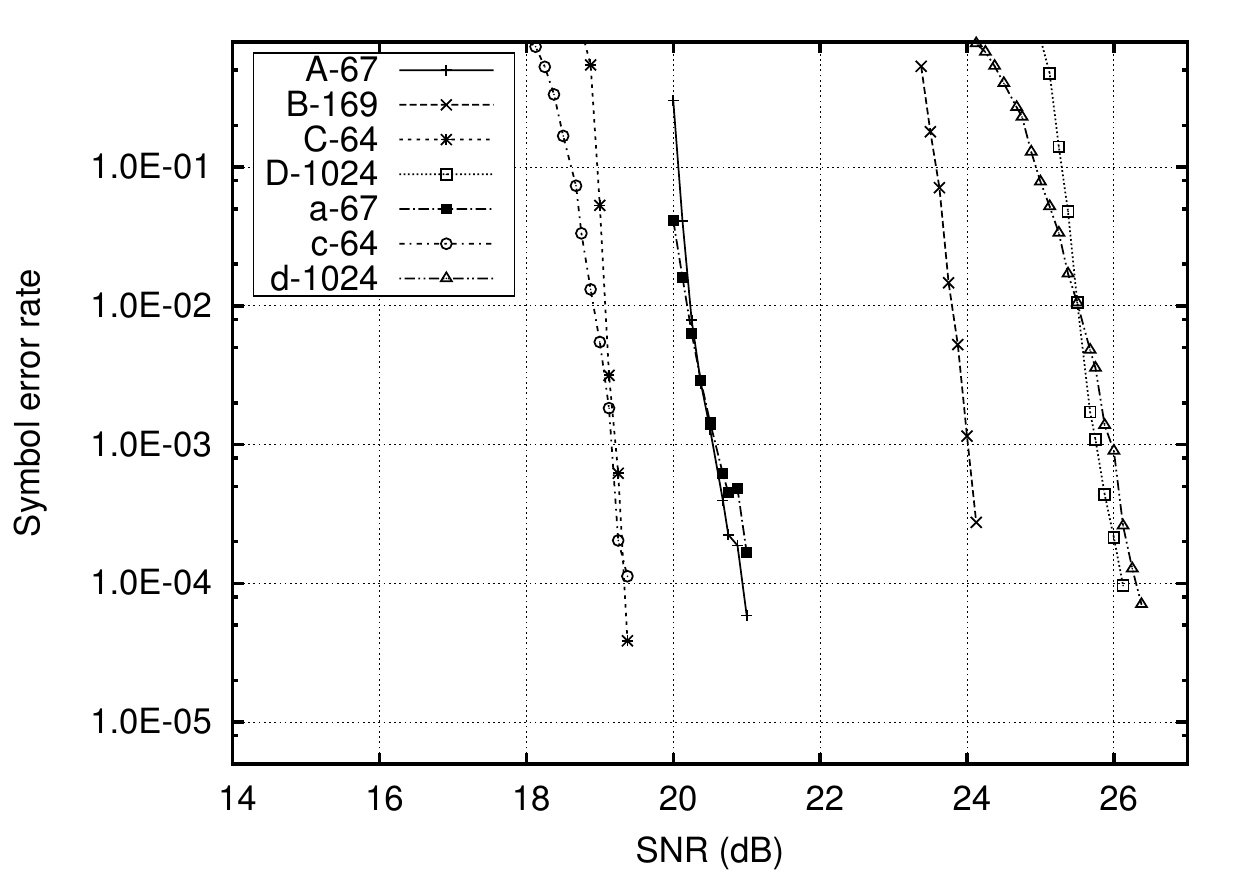}
  \caption{Results of symbol error rate versus relative SNR obtained
    simulating with different constellations and codeword lengths.  The constellations are indicated by
their size in the labels of the curves. With uppercase we
indicate long codes, $N=524288$ for the 169-QAM constellation and
$N=65536$ for the rest of long codes. The short codes, N=2048, are
labeled with lower case letters. The circular constellation and codes A-67 and a-67 where designed
for a target
$SNR=20.0$ dB, the rectangular B-169 for $SNR=23$ dB, the
rectangular C-64 and c-64 for $SNR=19$ dB, and the rectangular D-1024
and d-1024 for $SNR=25$ dB.}
\label{fig:6}
\end{figure}

In Fig. \ref{fig:6} we represent symbol error rate
versus SNR measured by counting errors between the input to the encoder
and the output from the decoder. The rectangular QAM constellation
uses independent coding in each PAM constellation. The constellation's size varies from 64
to 1024 points and the codeword length from 2048 to 524288. 
The codes were constructed with the criteria commented above
and the channel code rates were $R_c=0.9631$ and $R_c=0.9507$ for codes A-67 and a-67, respectively; $R_c=0.9242$ and $R_c=0.9801$
for codes C-64 and c-64 respectively; $R_c=0.7549$ and
$R_c=0.707$ for codes D-1024 and d-1024 respectively, and finally,
$R_c=0.9030$ for the code B-169. 

A first conclusion from the results shown in Fig. \ref{fig:6} is that
rectangular constellations with sizes $p^2$, $p$ prime, need very long codes in order to have
error rate figures similar to that obtained with the other constellations considered
in this paper. They use prime sized PAM constellations and therefore the polar
codes use the basic polar transformation \cite{arikan09}. The
simulations show that the curves of ordered Bhattacharyya parameters
versus index number, corresponding to PAM constellations with prime
size, do not have a fast fall when the constellation size is larger
than 7. This problem does not appear with PAM
constellations with size $2^m$, $m\in\mathbb{N}$, where the extended
polar transformation \eqref{eq:1} is used. In fact, it is possible to
construct very large QAM constellations, as shown in curves
D-1024 and d-1024, working with PAM constellations of size $2^m$,
$m\in\mathbb{N}$,  coding with the extended polar transformation
\eqref{eq:1}.

\begin{figure}[ht]
  \centering
  \includegraphics[width=9cm]{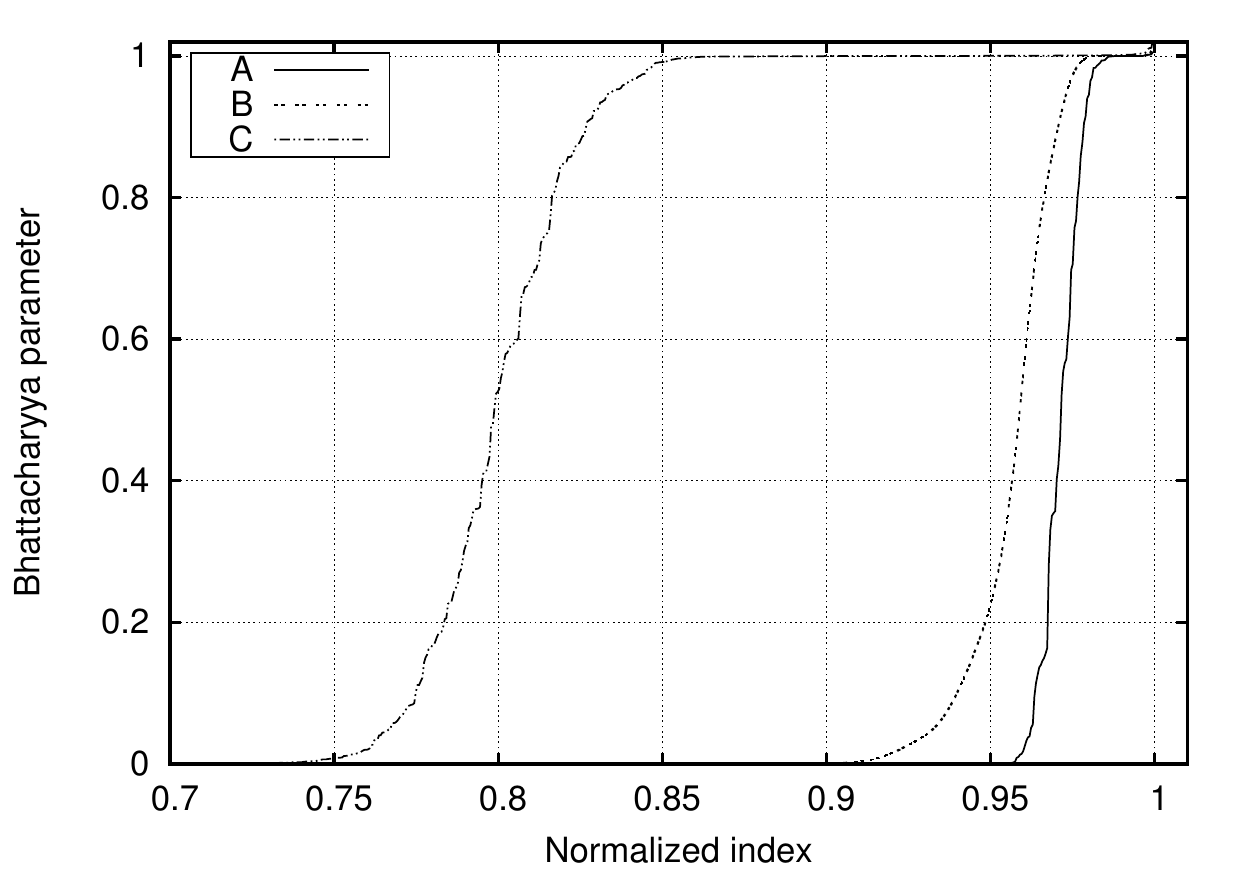}
  \caption{Ordered Bhattacharyya parameters versus the
    indices normalized by the codeword length of three of the codes
    considered in Fig. \ref{fig:6} . The curve A is for the
    polar code with a circular constellation with 67 points and codeword
    length $N=2048$. The curve B is for the polar code used with a
    rectangular constellation of 169 points and codeword length
    $N=524288$. In curve C the codeword length is $N=2048$ with a rectangular 1024-QAM constellation.} 
\label{fig:7}
\end{figure}
    
In Fig. \ref{fig:7} we compare the Bhattacharyya parameters of the
longest and the shortest polar codes of Fig. \ref{fig:6}. The curve
A in Fig. \ref{fig:7}, corresponding to the short polar code with
circular constellation, has a faster fall than that of the longer
codes. The shape of the curves B and C is similar in the region of
low Bhattacharyya parameters. This is interesting and shows that with
rectangular constellations and relatively short codes, and coding with the extended
transformation \eqref{eq:1}, it is possible to obtain symbol error
figures similar to the ones obtained with rectangular constellations
and the basic transformation, but much longer codeword lengths.

\subsection{Comparison with Other Techniques}

We have compared the direct q-ary polar approach presented in this work with
the two main coding techniques for Gaussian channels with multilevel
signals: Multilevel coding (MLC) \cite{wachsmann99} and
bit-interleaved coded modulation (BICM) \cite{caire98}.  We have used
\cite{seidl13} as a reference for comparison because it gathers
results obtained with multilevel polar coding (MLPC) and BICM using binary polar codes. Therefore
it is possible to compare the results obtained using binary polar
codes for MLPC or BICM with the more direct approach here
presented. Furthermore, there is a comparison in \cite{seidl13} 
between MLPC and the BICM scheme of the DVB-T2
standard \cite{dvbt2}. This comparison is an indication of how
adequate are polar codes for real applications.

Three signal sets are considered with sizes M=4, M=16, and M=256. The
codeword length of the concatenated LDPC
and BCH code of the DVB-T2 standard
is N=64800. In \cite{seidl13} the overall length of the MLPC
used in the comparisons was N=65536. The codeword lengths of the
component binary polar codes were: N=32768 for M=4, N=16384 for M=16 and
N=8192 for  M=256. We have chosen the same codeword lengths for the
q-ary polar codes in order to make a fair comparison. The maximum word
error rate (WER) was $\text{WER}_{\text{max}}$=$10^{-7}$. Labeling is
important for codes with MLC in order to obtain good codes
\cite{wachsmann99,seidl13}. Set-partitioning labeling was the choice for the
MLPC used for comparison with the
BICM scheme of DVB-T2 \cite{seidl13}. However, no optimization of labeling is
needed for q-ary polar codes. 

\begin{figure}[ht]
  \centering
  \includegraphics[width=9cm]{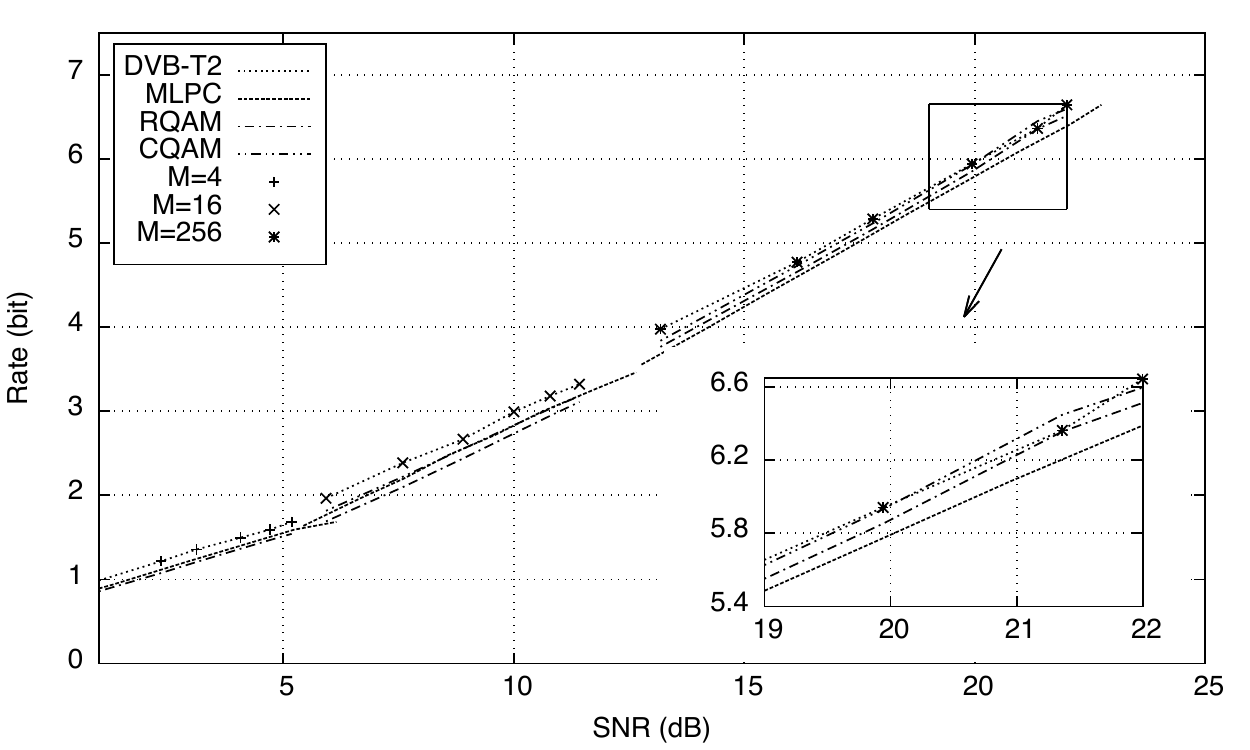}
  \caption{Comparison of multilevel polar codes, BICM codes of DVB-T2
    and q-ary polar codes.} 
\label{fig:9}
\end{figure}

In Fig. \ref{fig:9} we present the results of the comparison. In the
x-axis the SNR and in the y-axis the rate in bits per symbol. The
data of the MLPC  and the BICM of DVB-T2 are from
\cite[Fig.~12]{seidl13}. The rates of the MLPC are obtained from
density evolution (DE). However, it is a difficult task to implement
DE with q-ary polar codes. The low value of WER required by DVB-T2
makes difficult to obtain information of rate from
counting errors in the simulations. The bound \eqref{eq:20}  has been
used to obtain the information set. If the information set $\cal{A}$
is formed with the set of indices such that $(q-1)\sum_{i\in \mathcal{A}}Z(U_i|Y_1^N,U_1^{i-1})\le
10^{-7}$ then WER$\le 10^{-7}$. The results from \cite {seidl13} come
from DE, they are approximate; however, the q-ary polar code results
are obtained from a worst case scenario. In Fig. \ref{fig:9} we have
included the rectangular shaped constellation, labelled as RQAM in the
curves, and the circular shaped constellation, label CQAM. One q-ary
polar code with all the constellation points has been used with the
rectangular QAM constellations. The curve
with label DVB-T2 indicates the reference values for DVB-T2 and the
curve with label MLPC  corresponds to the MLPC scheme with data from
\cite{seidl13}. The marks on the DVB-T2 curves indicate the size of
the signal sets. We have not put marks on the other curves because it
is easy to identify them as they are close to the DVB-T2 ones.

The results in Fig. \ref{fig:9} show that for small signal sets, M=4
and M=16, MLPC is  slightly better than q-ary polar coding with
rectangular shaped constellation, curves with label RQAM. For M=16 the curves of MLPC and
q-ary polar coding with circular constellations, curve CQAM,
coincide. Both coding systems, MLPC and q-ary polar coding, have
curves below that of the DVB-T2. However, for large signal sets,
M=256, q-ary polar codes with both rectangular and circular
shaped constellations have curves above the MLPC one. Furthermore,
with q-ary polar coding and circular shaped constellation it is
possible to approach the DVB-T2 objectives working with high SNR.

We have not made comparisons with BICM schemes with polar codes as,
according to \cite{seidl13}, they operate further from capacity
than the MLPC schemes. Using the results in \cite{wachsmann99} we have compared a MLC with a turbo code as component
code against q-ary polar codes. The TC codeword length was
N=20000. Other conditions were: SNR=16.865 dB, bit error rate (BER)
=$10^{-5}$, signal set size M=64. The rate was \cite{wachsmann99}
R=5 bits.  With a q-ary polar code and
rectangular shaped constellation, codeword length N=16384, with same
SNR, M, and BER, the rate was R=5.00 bits. With a circular shaped
constellation the rate was R=5.05 bits. With a q-ary polar code,
rectangular shaped constellation, and N=32768, the rate was R=5.04 bits.

The complexity of  q-ary polar decoding is quadratic in the alphabet
size. The decoding complexity of MLPC is linear in the
alphabet size. However, q-ary polar decoding is more adequate for a
parallel implementation and can be implemented with the current
technology. The comparison of the complexity of q-ary polar coding with MLC with
an arbitrary component code is more involved as it depends in the
particular component code.

A summary of the comparisons done with other codes is the following:
for 1D signal sets MLC and MLPC are slightly better, i.e., closer to
capacity, than q-ary polar codes. For 2D signal sets with small size,
same results as for 1D case. However, for 2D signal sets with size
64 and over, q-ary polar codes are better than MLC or MLPC with both circular and
rectangular constellations.

%\cite[Eq.~75~and~76]{arikan09

\section{Conclusions}
\label{sec:conclusions}

In this paper we have presented a direct approach to the problem of
q-ary polar encoding and decoding for sources and channels. The encoding
process is very efficient. We express the LR for decoding in a recursive
equation by means of a LR vector, allowing the
implementation of  SC decoding  in a way that is similar to the
binary case. The decoding algorithm is quadratic in the size of the field
used for coding, however, the LR vector softens the complexity by
a factor of $1/2$ and many of the more complex operations can be parallelized. The Bhattacharyya parameters can be put as a
function of the decoding LR allowing the construction of the code
using Monte Carlo, as with the binary case. We have done numerical experiments of direct q-ary
polar coding for sources and channels with alphabets and constellations
of sizes from 5 to 1024 and codeword lengths between 2048 and 524288. For the Gaussian channel we have used
rectangular QAM constellations and optimal circular constellations. We have also compared the direct q-ary polar coding with
other multilevel schemes for channel coding. The results show that direct q-ary polar coding
is closer to the theoretical limit than other alternatives when the
alphabet size is large. Therefore it is possible to apply direct
q-ary polar coding to real scenarios with relatively short codes.

\pagebreak

\section*{Appendix A}
{\sc Proof of Claim 1}

The extended polar transformation \eqref{eq:1} is invertible and so is the polar transform
from $X_1^N$ to $U_1^N$ shown in Fig. \ref{fig:1}. Therefore we have,
%{\frac{N}{2}}
\begin{equation}
  \label{eq:2}
\begin{split}
  {P_{N}}_{U_1^{N}|Y_1^{N}}&(u_1^{N}|y_1^{N})\\
 &= {P_{{\frac{N}{2}}}}_{S_1^{{\frac{N}{2}}}|Y_1^{{\frac{N}{2}}}}(s_1^{{\frac{N}{2}}}|y_1^{{\frac{N}{2}}})
  {P_{{\frac{N}{2}}}}_{S_{{\frac{N}{2}}+1}^{N}|Y_{{\frac{N}{2}}+1}^{N}}(s_{{\frac{N}{2}}+1}^{N}|y_{{\frac{N}{2}}+1}^{N}).
\end{split}
\end{equation}
In order to alleviate the notation,
hereafter we omit the subscripts from the probabilities when they are
obvious from the context.

From \eqref{eq:2} and Fig. \ref{fig:1} we have,
\begin{equation}
  \label{eq:3}
   P_{N}(u_1^{N}|y_1^{N}) = P_{N}(u_{1,o}^{N}-\alpha u_{1,e}^{N}|y_1^{{\frac{N}{2}}})
   P_{N}(u_{1,e}^{N}|y_{{\frac{N}{2}}+1}^{N}) ,
\end{equation}
where the subscripts $1,o$, or $1,e$, indicate the odd, or even, indices,
respectively, of the vector and the minus sign refers to the sum of
the additive inverse element in $\mathbb{F}_q$. 

We use \eqref{eq:3} to obtain,
\begin{equation}
  \label{eq:4}
  \begin{aligned}
    P_{N}&(u_1^{2i-1}|y_1^{N})\\
&=\sum_{u_{2i,o}^{N}}\sum_{u_{2i,e}^{N}}P_{{\frac{N}{2}}}(u_{1,o}^{N}-\alpha u_{1,e}^{N}|y_1^{{\frac{N}{2}}})
   P_{{\frac{N}{2}}}(u_{1,e}^{N}|y_{{\frac{N}{2}}+1}^{N})\\
&=\sum_{u_{2i}}\sum_{u_{2i+1,e}^{N}} P_{{\frac{N}{2}}}(u_{1,e}^{N}|y_{{\frac{N}{2}}+1}^{N}) 
     \underbrace{\sum_{u_{2i+1,o}^{N}}
       P_{{\frac{N}{2}}}(u_{1,o}^{N}-\alpha u_{1,e}^{N}|y_1^{{\frac{N}{2}}})}_{P_{{\frac{N}{2}}}(u_{1,o}^{2i}-\alpha u_{1,e}^{2i}|y_1^{{\frac{N}{2}}})}\\
&=\sum_{u_{2i}}P_{{\frac{N}{2}}}(u_{1,o}^{2i}-\alpha u_{1,e}^{2i}|y_1^{{\frac{N}{2}}})
    \underbrace{
      \sum_{u_{2i+1,e}^{N}}P_{{\frac{N}{2}}}(u_{1,e}^{N}|y_{{\frac{N}{2}}+1}^{N})}_{P_{{\frac{N}{2}}}(u_{1,e}^{2i}|y_{{\frac{N}{2}}+1}^{N})}\\
&=\sum_{u_{2i}}P_{{\frac{N}{2}}}(u_{1,o}^{2i}-\alpha u_{1,e}^{2i}|y_1^{{\frac{N}{2}}}) P_{{\frac{N}{2}}}(u_{1,e}^{2i}|y_{{\frac{N}{2}}+1}^{N}).
  \end{aligned}
\end{equation}
Similarly,
\begin{equation}
  \label{eq:5}
  \begin{aligned}
     P_{N}&(u_1^{2i}|y_1^{N})\\
&=\sum_{u_{2i+1,e}^{N}}
   P_{{\frac{N}{2}}}(u_{1,e}^{N}|y_{{\frac{N}{2}}+1}^{N})
   \underbrace{\sum_{u_{2i+1,o}^{N}}P_{{\frac{N}{2}}}(u_{1,o}^{N}-\alpha u_{1,e}^{N}|y_1^{{\frac{N}{2}}})}_{P_{{\frac{N}{2}}}(u_{1,o}^{2i}-\alpha u_{1,e}^{2i}|y_1^{{\frac{N}{2}}})}\\
&=P_{{\frac{N}{2}}}(u_{1,o}^{2i}-\alpha u_{1,e}^{2i}|y_1^{{\frac{N}{2}}}) P_{{\frac{N}{2}}}(u_{1,e}^{2i}|y_{{\frac{N}{2}}+1}^{N}).
  \end{aligned}
\end{equation}

The conditional probability of the symbol $u_{2i-1}$ is obtained from
\eqref{eq:3}
\begin{equation}
  \label{eq:6}
  \begin{aligned}
     P_{N}^{(2i-1)}&(u_{2i-1}|y_1^{N},u_1^{2i-2})\\
&=\frac{1}{
       P_{N}(u_1^{2i-2}|y_1^{N})} P_{N}(u_1^{2i-1}|y_1^{N})\\
&=\frac{\sum_{u_{2i}}P_{{\frac{N}{2}}}(u_{1,o}^{2i}-\alpha u_{1,e}^{2i}|y_1^{{\frac{N}{2}}})
  P_{{\frac{N}{2}}}(u_{1,e}^{2i}|y_{{\frac{N}{2}}+1}^{N})}%
{P_{N}(u_1^{2i-2}|y_1^{N})},
  \end{aligned}
\end{equation}
where we have used the superscript $(2i-1)$ and the subscript $N$ to
highlight the similarity of this probability with the coordinate
channel defined in \cite{arikan09}.
  Using the fact,
  \begin{equation}
    \label{eq:7}
    \begin{split}
      P_{{\frac{N}{2}}}&(u_{1,o}^{2i}-\alpha u_{1,e}^{2i}|y_1^{{\frac{N}{2}}})\\
&=P_{{\frac{N}{2}}}^{(2i-1)}(u_{2i-1}-\alpha u_{2i}|y_1^{{\frac{N}{2}}},u_{1,0}^{2i-2}-\alpha u_{1,e}^{2i-2})\\
    &\qquad \times P_{{\frac{N}{2}}}(u_{1,o}^{2i-2}-\alpha u_{1,e}^{2i-2}|y_1^{{\frac{N}{2}}})\\
      P_{{\frac{N}{2}}}&(u_{1,e}^{2i}|y_{{\frac{N}{2}}+1}^{N})\\
&=P_{{\frac{N}{2}}}^{(2i)}(u_{2i}|y_{{\frac{N}{2}}+1}^{N},u_{1,e}^{2i-2}) P_{{\frac{N}{2}}}(u_{1,e}^{2i-2}|y_{{\frac{N}{2}}+1}^{N}),
    \end{split}
  \end{equation}
we can obtain from \eqref{eq:6} the following expression for
$P_{N}^{(2i-1)}(u_{2i-1}|y_1^{N},u_1^{2i-2})$,
  \begin{equation}
    \label{eq:8}
    \begin{aligned}
      P_{N}^{(2i-1)}&(u_{2i-1}|y_1^{N},u_1^{2i-2})\\
&=\frac{%
P_{{\frac{N}{2}}}(u_{1,o}^{2i-2}-\alpha u_{1,e}^{2i-2}|y_1^{{\frac{N}{2}}}) P_{{\frac{N}{2}}}(u_{1,e}^{2i-2}|y_{{\frac{N}{2}}+1}^{N})%
}{%
P_{N}(u_1^{2i-2}|y_1^{N})
}\\
&\ \ \times \sum_{u_{2i}}
P_{{\frac{N}{2}}}^{(i)}(u_{2i-1}-\alpha u_{2i}|y_1^{{\frac{N}{2}}},u_{1,0}^{2i-2}-\alpha u_{1,e}^{2i-2})\\
&\qquad \qquad \times P_{{\frac{N}{2}}}^{(i)}(u_{2i}|y_{{\frac{N}{2}}+1}^{N},u_{1,e}^{2i-2}).
    \end{aligned}
  \end{equation}
Using similar arguments we have,
\begin{equation}
  \label{eq:9}
  \begin{aligned}
         P_{N}^{(2i)}&(u_{2i}|y_1^{N},u_1^{2i-1})\\
&=\frac{%
P_{{\frac{N}{2}}}(u_{1,o}^{2i-2}-\alpha u_{1,e}^{2i-2}|y_1^{{\frac{N}{2}}}) P_{{\frac{N}{2}}}(u_{1,e}^{2i-2}|y_{{\frac{N}{2}}+1}^{N})%
}{%
P_{N}(u_1^{2i-1}|y_1^{N})
}\\
&\quad \times 
P_{{\frac{N}{2}}}^{(i)}(u_{2i-1}-\alpha u_{2i}|y_1^{{\frac{N}{2}}},u_{1,0}^{2i-2}-\alpha u_{1,e}^{2i-2})\\
&\quad \times P_{{\frac{N}{2}}}^{(i)}(u_{2i}|y_{{\frac{N}{2}}+1}^{N},u_{1,e}^{2i-2}).
  \end{aligned}
\end{equation}

From \eqref{eq:8} we can express
$L_{N}^{(2i-1)}(u|y_1^{N},u_1^{2i-2})$ as,
\begin{equation}
  \label{eq:12}
  \begin{aligned}
    L_{N}^{(2i-1)}&(u|y_1^{N},u_1^{2i-2})\\
&=\frac{%
P_{N}^{(2i-1)}(u|y_1^{N},u_1^{2i-2})}%
{P_{N}^{(2i-1)}(0|y_1^{N},u_1^{2i-2})}=\\
&\mkern-74mu\frac{%
\sum_{u_{2i}}
P_{{\frac{N}{2}}}^{(i)}(u-\alpha u_{2i}|y_1^{{\frac{N}{2}}},u_{1,0}^{2i-2}-\alpha u_{1,e}^{2i-2})%
P_{{\frac{N}{2}}}^{(i)}(u_{2i}|y_{{\frac{N}{2}}+1}^{N},u_{1,e}^{2i-2})
}{%
\sum_{u_{2i}}
P_{{\frac{N}{2}}}^{(i)}(-\alpha u_{2i}|y_1^{{\frac{N}{2}}},u_{1,0}^{2i-2}-\alpha u_{1,e}^{2i-2})%
P_{{\frac{N}{2}}}^{(i)}(u_{2i}|y_{{\frac{N}{2}}+1}^{N},u_{1,e}^{2i-2}), 
}\\
&\qquad  \ u\in {\mathbb{F}_q}, u\ne 0.
  \end{aligned}
\end{equation}
After dividing the numerator and denominator of the last term of
\eqref{eq:12} by 
\label{sec:appendix-2}
\[P_{{\frac{N}{2}}}^{(i)}(0|y_1^{{\frac{N}{2}}},u_{1,0}^{2i-2}-\alpha u_{1,e}^{2i-2})%
P_{{\frac{N}{2}}}^{(i)}(0|y_{{\frac{N}{2}}+1}^{N},u_{1,e}^{2i-2}) \] we obtain the first
result of \eqref{eq:11}. The second result is obtained in a similar
way. First from \eqref{eq:9} we have,

\begin{equation}
  \label{eq:13}
  \begin{aligned}
    L_{N}^{(2i)}&(u|y_1^{N},u_1^{2i-2},u_{2i-1})\\
&=\frac{%
P_{N}^{(2i)}(u|y_1^{N},u_1^{2i-1})}%
{P_{N}^{(2i)}(0|y_1^{N},u_1^{2i-1})}\\
&\mkern-72mu=\frac{%
P_{{\frac{N}{2}}}^{(i)}(u_{2i-1}-\alpha u|y_1^{{\frac{N}{2}}},u_{1,0}^{2i-2}-\alpha u_{1,e}^{2i-2})%
P_{{\frac{N}{2}}}^{(i)}(u|y_{{\frac{N}{2}}+1}^{N},u_{1,e}^{2i-2})
}{%
P_{{\frac{N}{2}}}^{(i)}(u_{2i-1}|y_1^{{\frac{N}{2}}},u_{1,0}^{2i-2}-\alpha u_{1,e}^{2i-2})%
P_{{\frac{N}{2}}}^{(i)}(0|y_{{\frac{N}{2}}+1}^{N},u_{1,e}^{2i-2})
}\\
&\qquad \ u\in {\mathbb{F}_q}, u\ne 0.
  \end{aligned}
\end{equation}
We obtain the second result in \eqref{eq:11} by dividing the numerator
and denominator of \eqref{eq:13} by 
\[P_{{\frac{N}{2}}}^{(i)}(0|y_1^{{\frac{N}{2}}},u_{1,0}^{2i-2}-\alpha u_{1,e}^{2i-2})\]

\section*{Appendix B}

{\sc Proof of Claim 2}

The Bhattacharyya parameter associated with  $P^{(1)}$ is
\begin{equation}
  \label{eq:a2e1}
  \begin{aligned}
        Z(&U_i|{Y_1}_1^N,U_1^{i-1})=\frac{1}{q-1} \sum_{\substack{u_i,u_i^\prime\\
        u_i\ne u_i^\prime}} \sum_{u_1^{i-1}} \sum_{{y_1}_1^N}\\
&\quad\sqrt{
      {P^{(1)}}_N^{(i)}(u_i,u_1^{i-1},{y_1}_1^N)
      {P^{(1)}}_N^{(i)}(u_i^\prime,u_1^{i-1}, {y_1}_1^N)}\\
&=\frac{1}{q-1}  \sum_{\substack{u_i,u_i^\prime\\
        u_i\ne u_i^\prime}}\sum_{u_1^{i-1}}\sum_{{y_1}_1^N}\\
    &\ \sqrt{ \sum_{u_{i+1}^N} {P^{(1)}}_N ({y_1}_1^N| u_i,u_1^{i-1}, u_{i+1}^N ) {P}_N (u_i,u_1^{i-1},u_{i+1}^N)}\\
     &\ \times \sqrt{\sum_{u_{i+1}^N} {P^{(1)}}_N ({y_1}_1^N| u_i^\prime,u_1^{i-1}, u_{i+1}^N ) {P}_N (u_i^\prime,u_1^{i-1},u_{i+1}^N)}.
  \end{aligned}
\end{equation}

Matrix $G_N$ is invertible. We call $S_{1}$, $S_i$, $S_{i+1}$ the
linear subspaces spanned by the rows $1$ to $i-1$, $i$, and $i+1$ to
$N$, respectively, of $G_N^{-1}$. There is a one-to-one correspondence
between $(u_{1}^{i-1},u_i, u_{i+1}^N)$ and
$\mathbf{x}_{u_i}+\mathbf{x}_{1}+\mathbf{x}_{2}$, where
$\mathbf{x}_{1}=(u_1^{i-1},0,0_{i+1}^N) G_N^{-1}$,
$\mathbf{x}_{u_{i}}=(0,u_i,0_{i+1}^N) G_N^{-1}$, and
$\mathbf{x}_{2}=(0_1^{i-1},0,u_{i+1}^{N}) G_N^{-1}$. Therefore Eq.
\eqref{eq:a2e1} is
%\displaybreak[0]

%\begin{equation}
\begin{align}
        Z(&U_i|{Y_1}_1^N,U_1^{i-1})=\frac{1}{q-1}%%
            \sum_{\substack{\mathbf{x}_{u_i},\mathbf{x}_{u_i^\prime}\in S_i\\
       \mathbf{x}_{u_i} \ne \mathbf{x}_{u_i^\prime}}
    }\sum_{\mathbf{x}_{1}\in S_1}\sum_{{y_1}_1^N}\notag \\
&\ \sqrt{ \sum_{\mathbf{x}_{2}\in S_{i+1}}  {P^{(1)}}_N({y_1}_1^N|\mathbf{x}_{u_i}+\mathbf{x}_{1}+\mathbf{x}_{2}) P_N(\mathbf{x}_{u_i}+\mathbf{x}_{1}+\mathbf{x}_{2})}\notag\\
&\times \sqrt{\sum_{\mathbf{x}_{2}\in S_{i+1}}
  {P^{(1)}}_N({y_1}_1^N|\mathbf{x}_{u_i^\prime}+\mathbf{x}_{1}+\mathbf{x}_{2})
  P_N(\mathbf{x}_{u_i^\prime}+\mathbf{x}_{1}+\mathbf{x}_{2})}\notag \displaybreak[0]\\
&=\frac{1}{q-1}
   \sum_{\substack{\mathbf{x}_{u_i},\mathbf{x}_{u_i^\prime}\in S_i\\
       \mathbf{x}_{u_i} \ne \mathbf{x}_{u_i^\prime}}  }\sum_{\mathbf{x}_{1}\in S_1}\sum_{{y_1}_1^N}\notag\\
&\Big( \sum_{\mathbf{x}_{2}\in S_{i+1}} P_N(\mathbf{x}_{u_i}+\mathbf{x}_{1}+\mathbf{x}_{2})\notag\\
  &\qquad \prod_{j=1}^N {P^{(1)}}({y_1}_j|{\mathbf{x}_{u_i}}_j+{\mathbf{x}_{1}}_j+{\mathbf{x}_{2}}_j) \Big)^{1/2}\notag\\
&\times \Big( \sum_{\mathbf{x}_{2}\in S_{i+1}} P_N(\mathbf{x}_{u_i^\prime}+\mathbf{x}_{1}+\mathbf{x}_{2}) \notag\\
   &\qquad \prod_{j=1}^N
  {P^{(1)}}({y_1}_j|{\mathbf{x}_{u_i^\prime}}_j+{\mathbf{x}_{1}}_j+{\mathbf{x}_{2}}_j)
  \Big)^{1/2},
\label{eq:a2e2}
\end{align}
%\end{equation}
where we have used the fact that the channel is memoryless and
${\mathbf{x}_a}_j$ is the $j$-th component of ${\mathbf{x}_a}$.
  
By hypothesis $P^{(1)}\preceq P^{(2)}$. From \eqref{eq:deg_chan}
$P^{(1)}$ is a function of $P^{(2)}$ and a channel $P$. Using this
fact in \eqref{eq:a2e2} we have,
\begin{equation}
  \label{eq:a2e3}
  \begin{aligned}
        Z(&U_i|{Y_1}_1^N,U_1^{i-1})=\frac{1}{q-1}
    \sum_{\substack{\mathbf{x}_{u_i},\mathbf{x}_{u_i^\prime}\in S_i\\
       \mathbf{x}_{u_i} \ne \mathbf{x}_{u_i^\prime}}  }\sum_{\mathbf{x}_{1}\in S_1}\sum_{{y_1}_1^N}\\
&\ \Big( \sum_{\mathbf{x}_{2}\in S_{i+1}}
P_N(\mathbf{x}_{u_i}+\mathbf{x}_{1}+\mathbf{x}_{2}) \\
   &\quad \prod_{j=1}^N \sum_{{y_2}_j} {P^{(2)}}({y_2}_j|{\mathbf{x}_{u_i}}_j+{\mathbf{x}_{1}}_j+{\mathbf{x}_{2}}_j) P({y_1}_j|{y_2}_j)\Big)^{1/2}\\
&\ \times\Big( \sum_{\mathbf{x}_{2}\in S_{i+1}} P_N(\mathbf{x}_{u_i^\prime}+\mathbf{x}_{1}+\mathbf{x}_{2}) \\
   &\quad \prod_{j=1}^N \sum_{{y_2}_j}
   {P^{(2)}}({y_2}_j|{\mathbf{x}_{u_i^\prime}}_j+{\mathbf{x}_{1}}_j+{\mathbf{x}_{2}}_j)
   P({y_1}_j|{y_2}_j)\Big)^{1/2}.\\
\end{aligned}
\end{equation}

By expanding the products in \eqref{eq:a2e3} we have,
%\begin{equation}
\begin{align}
  %\label{eq:a2e4}
  %\begin{aligned}
 Z(&U_i|{Y_1}_1^N,U_1^{i-1})=\frac{1}{q-1}
   \sum_{\substack{\mathbf{x}_{u_i},\mathbf{x}_{u_i^\prime}\in S_i\\
       \mathbf{x}_{u_i} \ne \mathbf{x}_{u_i^\prime}}  }\sum_{\mathbf{x}_{1}\in S_1}\sum_{{y_1}_1^N}\notag\\
&\quad \Big( \sum_{\mathbf{x}_{2}\in S_{i+1}} \sum_{{y_2}_1}\cdots \sum_{{y_2}_N} P_N(\mathbf{x}_{u_i}+\mathbf{x}_{1}+\mathbf{x}_{2}) \notag \\
   &\qquad \prod_{j=1}^N  {P^{(2)}}({y_2}_j|{\mathbf{x}_{u_i}}_j+{\mathbf{x}_{1}}_j+{\mathbf{x}_{2}}_j) P({y_1}_j|{y_2}_j)\Big)^{1/2}\displaybreak[0]\notag\\
&\ \times \Big( \sum_{\mathbf{x}_{2}\in S_{i+1}} \sum_{{y_2}_1}\cdots \sum_{{y_2}_N} P_N(\mathbf{x}_{u_i^\prime}+\mathbf{x}_{1}+\mathbf{x}_{2}) \notag\\
   &\qquad \prod_{j=1}^N  {P^{(2)}}({y_2}_j|{\mathbf{x}_{u_i^\prime}}_j+{\mathbf{x}_{1}}_j+{\mathbf{x}_{2}}_j) P({y_1}_j|{y_2}_j)\Big)^{1/2}\displaybreak[0]\notag\\
&\overset{(a)}{\ge} \frac{1}{q-1}
   \quad\times \sum_{\substack{\mathbf{x}_{u_i},\mathbf{x}_{u_i^\prime}\in S_i\\
       \mathbf{x}_{u_i} \ne \mathbf{x}_{u_i^\prime}}  }\sum_{\mathbf{x}_{1}\in S_1}\sum_{{y_1}_1^N} \sum_{{y_2}_1}\cdots \sum_{{y_2}_N}\notag\\
&\ \Big( \sum_{\mathbf{x}_{2}\in S_{i+1}} P_N(\mathbf{x}_{u_i}+\mathbf{x}_{1}+\mathbf{x}_{2}) \notag\\
   &\quad \prod_{j=1}^N {P^{(2)}}({y_2}_j|{\mathbf{x}_{u_i}}_j+{\mathbf{x}_{1}}_j+{\mathbf{x}_{2}}_j) P({y_1}_j|{y_2}_j)\Big)^{1/2}\displaybreak[0]\notag\\
&\times \Big( \sum_{\mathbf{x}_{2}\in S_{i+1}} P_N(\mathbf{x}_{u_i^\prime}+\mathbf{x}_{1}+\mathbf{x}_{2}) \notag\\
   &\quad \prod_{j=1}^N {P^{(2)}}({y_2}_j|{\mathbf{x}_{u_i^\prime}}_j+{\mathbf{x}_{1}}_j+{\mathbf{x}_{2}}_j) P({y_1}_j|{y_2}_j)\Big)^{1/2},
\label{eq:a2e4}
%\end{aligned}
%\end{equation}
\end{align}
where $(a)$ follows from the Schwarz's inequality. The sum of the
factor $P({y_1}_j|{y_2}_j)$ over all possible values of ${y_1}_j$ is
one, and therefore \eqref{eq:a2e4} is
\begin{equation}
  \label{eq:a2e33}
  \begin{aligned}
        Z(&U_i|{Y_1}_1^N,U_1^{i-1})\\
&\ge\frac{1}{q-1}
   \sum_{\substack{\mathbf{x}_{u_i},\mathbf{x}_{u_i^\prime}\in S_i\\
       \mathbf{x}_{u_i} \ne \mathbf{x}_{u_i^\prime}}  }\sum_{\mathbf{x}_{1}\in S_1} \sum_{{y_2}_1}\cdots \sum_{{y_2}_N}\\
&\quad \Big( \sum_{\mathbf{x}_{2}\in S_{i+1}} P_N(\mathbf{x}_{u_i}+\mathbf{x}_{1}+\mathbf{x}_{2}) \\
   &\qquad \prod_{j=1}^N {P^{(2)}}({y_2}_j|{\mathbf{x}_{u_i}}_j+{\mathbf{x}_{1}}_j+{\mathbf{x}_{2}}_j) \Big)^{1/2}\\
&\quad \times \Big( \sum_{\mathbf{x}_{2}\in S_{i+1}} P_N(\mathbf{x}_{u_i^\prime}+\mathbf{x}_{1}+\mathbf{x}_{2}) \\
   &\qquad \prod_{j=1}^N {P^{(2)}}({y_2}_j|{\mathbf{x}_{u_i^\prime}}_j+{\mathbf{x}_{1}}_j+{\mathbf{x}_{2}}_j) \Big)^{1/2}\\
& =\frac{1}{q-1}
   \sum_{\substack{\mathbf{x}_{u_i},\mathbf{x}_{u_i^\prime}\in S_i\\
       \mathbf{x}_{u_i} \ne \mathbf{x}_{u_i^\prime}}  }\sum_{\mathbf{x}_{1}\in S_1} \sum_{{y_2}_1^N}\\
&\quad \Big(  {P^{(2)}}_N({y_2}_1^N,{\mathbf{x}_{u_i}}+{\mathbf{x}_{1}}) \Big)^{1/2}\\
&\quad \times \Big( \prod_{j=1}^N {P^{(2)}}_N({y_2}_1^N,{\mathbf{x}_{u_i^\prime}}+{\mathbf{x}_{1}}) \Big)^{1/2}\\
& =\frac{1}{q-1}
   \sum_{\substack{u_i,{u_i^\prime}\\
       u_i \ne u_i^\prime}}  \sum_{u_1^{i-1}} \sum_{{y_2}_1^N}\\
&\quad \Big(  {P^{(2)}}_N({y_2}_1^N,u_i, u_1^{i-1}) \Big)^{1/2}\\
&\quad\times \Big(  {P^{(2)}}_N({y_2}_1^N,u_i^\prime, u_1^{i-1}) \Big)^{1/2}\\
&=  Z(U_i|{Y_2}_1^N,U_1^{i-1}).
\end{aligned}
\end{equation}
Thus,
\begin{equation}
  \label{eq:app2_28}
Z(U_i|{Y_1}_1^N,U_1^{i-1})\ge Z(U_i|{Y_2}_1^N,U_1^{i-1}),
\end{equation}
which had to be proved.

\section{Acknowledgement}
\label{sec:acknowledgement}

We would like to thank Prof. P. M. Olmos and Prof. T. Koch for their valuable suggestions.

\bibliographystyle{IEEEtran}
\bibliography{art_mpolar}

\end{document}